\documentclass[letterpaper,12pt]{article}
\usepackage{setspace}
\usepackage{graphicx}                
\usepackage{float}                   
\usepackage{hyperref}                
\usepackage[margin=1in]{geometry}  
\usepackage[short]{datetime}         
\usepackage[labelfont=bf]{caption}   
\usepackage[bottom]{footmisc}        
\usepackage{lineno}
\usepackage{ragged2e}
\usepackage[protrusion=true,expansion=true]{microtype} 
\usepackage[T1]{fontenc}                               
\usepackage{lmodern}
\usepackage{subfig}

\usepackage[compact]{titlesec}
\titleformat*{\section}{\large\bfseries}
\titlespacing*{\section}{0pt}{1ex}{1ex}
\titleformat*{\subsection}{\normalsize\bfseries}
\titlespacing*{\subsection}{0pt}{0ex}{0ex}
\titleformat*{\subsubsection}{\normalsize\bfseries}
\titlespacing*{\subsubsection}{0pt}{0ex}{0ex}

\usepackage{amsfonts,amssymb,amsmath, mathtools}
\usepackage{xcolor, color, soul}
\usepackage{graphicx}
\usepackage{multicol}
\usepackage{url}
\usepackage{faktor}
\usepackage{ulem}
\usepackage{pdfpages}
\usepackage{soul}
\usepackage{hyperref}
\usepackage{natbib}
\usepackage{setspace}
\DeclareMathAlphabet\mathbfcal{OMS}{cmsy}{b}{n}

\begin{document}

\center{\Large Adjacency Matrix Decomposition Clustering for Human Activity Data}
\vspace{-.05in}
\center{ Martha Barnard$^1$*, Yingling Fan$^2$, Julian Wolfson$^1$}
\vspace{-.1in}
\center{\footnotesize $^1$Division of Biostatistics and Health Data Science, School of Public Health, University of Minnesota \\
$^2$Department of Urban and Regional Planning Area, Hubert H. Humphrey School of Public Affairs, University of Minnesota}

\justifying
\begin{abstract}
Mobile apps and wearable devices accurately and continuously measure human activity; patterns within this data can provide a wealth of information applicable to fields such as transportation and health. Despite the potential utility of this data, there has been limited development of analysis methods for sequences of daily activities. In this paper, we propose a novel clustering method and cluster evaluation metric for human activity data that leverages an adjacency matrix representation to cluster the data without the calculation of a distance matrix. Our technique is substantially faster than conventional methods based on computing pairwise distances via sequence alignment algorithms and also enhances interpretability of results. We compare our method to distance-based hierarchical  clustering and nTreeClus through simulation studies and an application to data collected by Daynamica, an app that turns sensor data into a daily summary of a user's activities. Among days that contain a large portion of time spent at home, our method distinguishes days that also contain multiple hours of travel or other activities, while both comparison methods fail to identify these patterns. We further identify which day patterns classified by our method are associated with higher concern for contracting COVID-19 with implications for public health messaging.


\end{abstract}
\vspace{-0.2in}
{{\it Keywords}: \small digital phenotyping, human activity, mobile health, dimension reduction, clustering } \\
\noindent\rule{2in}{0.4pt} \\
{\small * barna126@umn.edu} \\

\vspace*{\fill}
\noindent\rule{6.3in}{0.4pt} \\
\noindent{\small Data and code for this paper are available at \url{https://github.com/m-barnard/amdc_clustering}}

\pagebreak


\section{Introduction}

With the ubiquity of mobile apps and wearable devices, we can measure human behavior in more detail than ever before. Phone- and wearable-based sensors accurately and continuously measure human activity to provide valuable information in the context of transportation, health, and other fields \citep{boaro2021, beukenhorst2021, panda2021, Song2021}. \cite{Torous2016} described the use of this data as digital phenotyping: the ``moment-by-moment quantification of the individual-level human phenotype in-situ using data from smartphones and other personal digital devices'' which emphasizes the continuous, or intensively sampled, nature of the data. One type of this data, continuous location tracking of human activities and movements in space, has enabled researchers to explore connections beyond people's home neighborhoods to where and how people work, shop, spend leisure time, attend school, and travel. For example, this data has enabled researchers in environmental health to generate mobility-based measures of human exposure to specific environmental amenities (e.g., green space) and risks (e.g., air pollution). Furthermore, continuous measurements of human activity have enabled time use researchers to classify people's time use patterns based upon the timing, sequence, and interdependence of activities during the day, not just the total amount of time spent on each type of activity. \cite{Freedman2019} used this data to investigate the relationship between time use sequences and emotional well-being among older caregivers. Using the continuous time use data, they were able to identify five distinctive caregiving patterns through clustering.  These patterns were found to vary by gender and work status of the caregiver.

\begin{figure}
    \centering
    \includegraphics[scale = 0.5]{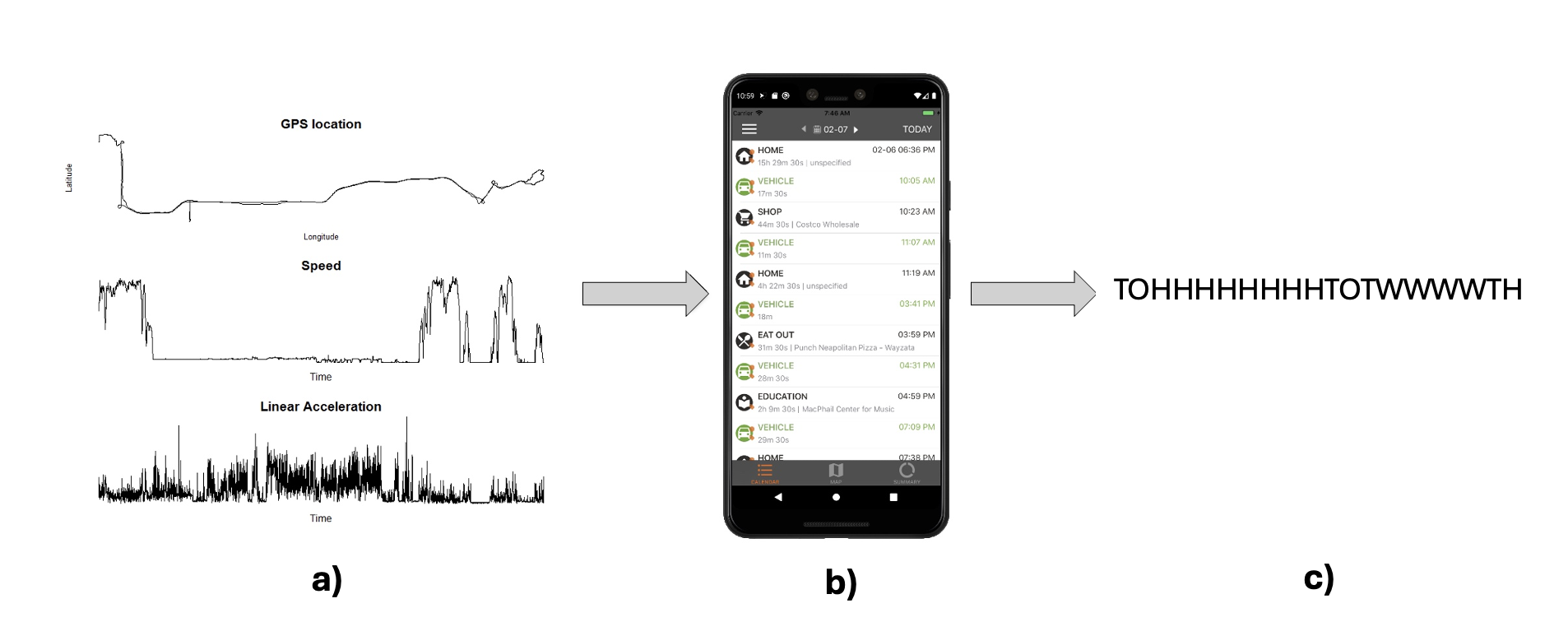}
    \caption{\footnotesize Sensor data, the Daynamica app calendar view of the data (episodic data), and the resulting sequence. The sequence can take the following states: H - Home, W - Work, T - Transport, and O - any other activity (Out).}
    \label{fig:daynamica}
\end{figure}

In this paper, we will focus on digital phenotype data recorded by Daynamica, a smartphone application that uses sensor-based activity tracking and ecological momentary assessment (EMA) to provide high resolution human activity data \citep{Fan2015}. Daynamica uses machine learning techniques to convert sensor data to a summary of a user's daily activities according to the major categories used by the American Time Use Survey \citep{atus2022}: home, work, school, eating out, shopping, leisure and recreation, and education. In addition, Daynamica detects the method of travel between activities, which is separated into travel by car, bus, train, bike, or walking. Individuals are asked, through the app, to correct any incorrectly converted data and to answer questions that supplement the sensor data, such as questions about emotional well-being or social activity. The resulting data provides a continuous 24-hour picture of how an individual spends their day in terms of activity states. Figures \ref{fig:daynamica}(a) and \ref{fig:daynamica}(b) provide an example of the sensor data and corresponding activity episode data within the Daynamica app.

Many studies collect a variety of mobile or wearable device data, including continuous (e.g., heart rate, physical acceleration \citep{karas2019, bolliger2020}) and categorical (e.g., location \citep{beukenhorst2021, panda2021}) outcomes. Previous studies that incorporate categorical mobile data, such as GPS activity data, have focused on univariate aggregate metrics, such as the amount of time spent at home, and then conducted standard statistical analyses on these metrics \citep{boaro2021, beukenhorst2021}. However, this simple aggregation does not take advantage of the richness of the data; while the data contains information about potentially many activity states, the time of day or week the states occur, and how the activity states relate to each other through time, a single metric can only capture one state at one time point. While this data could be modeled using complex multi-state models (e.g., competing risks, Markov models), these models tend to be computationally intensive and difficult to interpret \citep{piccarreta2019}. Clustering methods are an attractive option for analyzing this data as multiple data attributes can be taken into account when determining cluster assignment. In addition, clustering results from mobile data can be easily related to other more traditional data sources (e.g., demographic information, daily surveys) through standard statistical techniques. For the Daynamica data, we are specifically interested in how daily activities and time use are related to concern about contracting COVID-19.

\subsection{Categorical Time-series Clustering}
While there have been few clustering methods designed specifically for mobile data, there are a variety of methods for clustering categorical time-series data more broadly. Most of these methods represent the data as a sequence, which was made common through life course analysis \citep{aisenbrey2010}. To create the sequence, each individual's data is divided into temporally ordered, equal-sized time segments, where each segment is labeled by the state or category that occurs for the majority of the time segment. The result is an ordered character sequence, where each character represents a specific categorical state. Figure \ref{fig:daynamica}(c) depicts an example of a sequence created from Daynamica activity episode data. There have been two main approaches to clustering these sequences: model-based clustering and distance-based clustering. 

The first approach assumes the data follows some stochastic model (often a Markov model) and aims to cluster the data by estimating the parameters of the stochastic model within each cluster. \cite{pamminger_model-based_2010} and \cite{fruhwirth-schnatter_model-based_2010-1} both present methods that cluster sequences based on finite mixtures of time-homogenous Markov chains. While these methods are attractive for their ability to accommodate the full, multidimensional data structure, model and computational complexity increases quickly as the number of states increases due to separate model parameters for each state transition \citep{piccarreta2019}. Increasing model complexity also does not guarantee a good model fit; \cite{huang2023} found that stochastic models, even when complex, were incapable of simulating data that was similar to the real activity data used within the model. In addition, \cite{garcia-magarinos_framework_2015} found that the \cite{pamminger_model-based_2010} method performed poorly in simulations even when initializing the MCMC process with the true cluster parameters.

The use of distance-based methods has generally been more widespread, especially in fields such as bioinformatics. These methods define some (dis)similiarty metric in order to characterize the relationships between sequences. To compare sequences, researchers often use the sequence alignment method (SAM), where each time position in two sequences is compared to calculate a distance metric \citep{corpet1988, sankoff1983}. A common distance metric used is the Levenshtein distance \citep{Levenshtein1965, sankoff1983}, which is the number of changes (insertions, deletions, or substitutions) needed to make two sequences identical. Life course analysis researchers have developed a variety of methods using these common distance metrics, including identifying representive sequence patterns and k-mediod clustering of the sequence distance matrix \citep{ritschard2013, studer2021}. However, hierarchical clustering of the sequence distance matrix has been the most common sequence clustering method across a variety of disciplines \citep{corpet1988,Freedman2019, Song2021}. While distance-based methods are popular, calculating the distance matrix can become computationally prohibitive as the number of sequences and sequence length increase. Recent works that adapt standard distance-based clustering to a novel distance metric also tend to have heavy computational burden \citep{garcia-magarinos_framework_2015}. Computational requirements are of specific concern as digital phenotype data has the potential for determining close to real-time interventions, which require real-time data analyses \citep{onnela2021}. 

Due to the computational requirements of distance-based methods using full sequence alignment, there as been development of methods that cluster the data based on the distance between \textit{features} of the sequence. \cite{ghassempour_clustering_2014} and \cite{de_angelis_mining_2014} combine model- and distance-based methods by clustering data that has first been transformed using a hidden Markov model. \cite{lopez-oriona_hard_2023} propose clustering on features that quantify the serial dependence within biological (e.g., DNA) sequences. In addition, text analysis and genomic methods have been developed where sequences are compared through their $n-$grams or $k-$mers, respectively \citep{Blaisdell1986, Kondrak2005, Kantorovitz2007}. Clustering on $n-$grams or $k-$mers is more computationally efficient than clustering on the Levenshtein distance matrix and has had superior performance in text classification problems and phylogenetic tree creation \citep{Peng2003, Wen2014}. \cite{jahanshahi_ntreeclus_2022} propose nTreeClus, which extends these methods by combining $k$-mers with an autoregressive decision tree to cluster the data based on common patterns. Feature distance-based methods have been shown to perform well for biological sequences; however, their focus on identifying patterns may not be as successful for clustering human activity data where there are fewer frequent state changes.

\subsection{Outline of proposed method}
In this paper, we propose a novel clustering method for sequential human activity data. Our technique uses an adjacency matrix representation of the sequences, clusters the sequences via a decomposition of the adjacency matrices, and evaluates the optimal cluster number with a new metric that adapts the Calinski-Harabasz index in the context of adjacency matrices. The technique provides considerable computational benefits in comparison to hierarchical clustering and nTreeClus (10 to 145 times faster with Daynamica data). The adjacency matrix representation and further decomposition increases the interpretability of clustering results and the ability to weight specific parts of the adjacency matrix provides flexibility within the clustering procedure. We explore the behavior of our method in comparison to hierarchical clustering with full sequence alignment and nTreeClus through simulations. We demonstrate the performance of our method on Daynamica activity data in which we identify meaningful daily and weekly activity patterns and their relationship to COVID-19 concerns. The structure of the paper is as follows: in Section 2, we describe the steps within our clustering method. We present simulation methods in Section 3 and the application of our method to Daynamica data in Section 4. We discuss results and future work in Section 5.

\section{Methods}

\subsection{Notation}
Our data consists of $n$ sequences, $\{y_i\}, i = 1, \ldots, n$. Each sequence $y_i$ is of common length $l$ and takes the form $\{y_{i,1}, \ldots, y_{i,l}\}$, where each $y_{i,j} \in \mathcal{S}$, a set of $m$ unique state $s_1, \ldots, s_m$. In this sequence formulation, $y_{i,j} = s_u$ represents an amount of time, $\delta$, spent in the state $s_u$ where $\delta$ is the same for every $j = 1, \ldots, l$.

\subsection{Clustering Method}
While our technique can be applied to any categorical sequence data, we present the method in the context of daily human activity data. Briefly, our clustering approach proceeds as follows. Each step is described in greater detail below:

\begin{enumerate}
\item Form an adjacency matrix $\mathbfcal{T}_i$ (i.e., a matrix encoding transitions between each pair of states) for sequence $y_i$.
\item Concatenate as columns the vectorized versions of each adjacency matrix to form a matrix $\mathbfcal{M}$.
\item Obtain the singular value decomposition of the centered matrix $\mathbfcal M_c = \mathbf{U \Sigma V}^T$.
\item Select the first $h$ columns of $\mathbf{V}$, and apply a clustering algorithm to them.
\item Apply a novel cluster evaluation metric to evaluate cluster quality.
\end{enumerate}

\subsubsection{Adjacency Matrix and SVD}
\label{sec:defining_adj_mat}
    For each sequence $y_i$, we create an $m \times m$ adjacency matrix $\mathbfcal{T}_i$. The $(u,v)$ entry of $\mathbfcal{T}_i$ is the number of times the $s_u$ state precedes the $s_v$ state in $y_i$, i.e.,
\begin{equation*}
  \mathbfcal{T}_{i_{u,v}} = \sum_{j = 1}^{l-1} I[y_{i,j} = s_u, y_{i,j+1} = s_v]
\end{equation*}
    Then, for $\{y_i\}$ a corresponding set of adjacency matrices is calculated, $\{\mathbfcal{T}_{i}\}$. We vectorize each adjacency matrix and obtain the set $\{\mathbf{t}_{i}\}$ where $\mathbf{t}_{i} = \text{vec}(\mathbfcal{T}_{i})$ such that $\mathbf{t}_{i}$ has length $m^2$. Then, we construct an $m^2 \times n$ matrix where the $i$th column is $\mathbf{t}_{i}$, $\mathbfcal{M} = [\mathbf{t}_{1}, \ldots \mathbf{t}_{n}]$. We center $\mathbfcal{M}$, $\mathbfcal{M}_c = \mathbfcal{M} - (l-1)/m^2\mathbf{J}_{m^2 \times n}$. Then, we decompose $\mathbfcal{M}_c$ using singular value decomposition (SVD), $\mathbfcal{M}_c = \mathbf{U \Sigma V}^T$, where $\mathbf{U},\mathbf{V}$ are orthogonal matrices and $\mathbf{\Sigma}$ is a diagonal matrix.

The adjacency matrix $\mathbfcal T_i$ is a compressed representation of the full sequence $y_i$. For each adjacency matrix, $\mathbfcal{T}_i$, the diagonal of the matrix is the number of same state transitions, which represents the contiguous duration spent in a state throughout the day (in relation to $\delta$).
All other entries represent how many times during the day there are transitions between states. By definition, we have $\sum_{u,v} \mathbfcal{T}_{{i}_{u,v}} = l - 1$. Scaling the entries of the adjacency matrix as $\mathbfcal{T}_{{i}_{u,v}} / \sum_u \mathbfcal{T}_{{i}_{u,v}}$ yields the estimated first-order Markov transition probability matrix for $y_i$. The first-order Markov matrix representation standardizes the row sums to one, however since the row sums provide useful information in this context (total amount of time spent in a state throughout the day) we use the described adjacency matrix representation. We also note that the vectorized form of the adjacency matrix, $\mathbf{t}_{i}$, is the $2-$mer count vector for the sequence $y_i$.

\subsubsection{SVD Output and Clustering}

After performing SVD, we select a subset of the columns of $\mathbf{V}$ that correspond to the $h$ largest singular values, $\mathbf{V}_{1:h}$ (the first $h$ principal axes in principal component analysis (PCA)). Each row of $\mathbf{V}_{1:h}$ corresponds to a single day sequence, so we can perform a cluster analysis on $\mathbf{V}_{1:h}$ to obtain a clustering of the sequences. We are essentially clustering on the numerical vectors corresponding to a transformation of the adjacency matrix representation of the set of day sequences.  There are multiple potential ways to select $h$, such as using methods for selecting principal axes in PCA,  however, we suggest that $h$ be selected through cluster evaluation metrics, which will be demonstrated in Sections \ref{sec:simulation} and \ref{sec:data_app}. 

Existing multidimensional clustering algorithms can be used to cluster $\mathbf{V}_{1:h}$ (such as k-means, hierarchical, DBSCAN, etc.). It is beyond the scope of this paper to evaluate the behavior of the clustering algorithms in this application, however there has been work comparing these methods in general \citep{Rodriguez2019}. We use the k-means algorithm in applications of the method in Sections 3 and 4.

    \subsubsection{Cluster Evaluation}

    Suppose that, based on the procedure described above, sequences are clustered into $p$ clusters where cluster $C_k$ contains $n_k$ sequences, $k = 1, \ldots, p.$  For each cluster, we construct a mean adjacency matrix for the cluster, $\mathbfcal{T}_{(k)} = \frac{1}{n_k} \sum_{i:y_i \in C_k} \mathbfcal{T}_{i}.$ 
    Using SVD, we decompose the matrices $\mathbfcal{T}_{(k)} = \mathbf{U}_{(k)}\mathbf{\Sigma}_{(k)}\mathbf{V}^{T}_{(k)}$ for each $\mathbfcal{T}_{(k)}$ and $\mathbfcal{T}_{i} = \mathbf{U}_i\mathbf{\Sigma}_i\mathbf{V}_i^T$ for each $\mathbfcal{T}_{i}$. The within-cluster distance metric is based on the difference between each $\mathbfcal{T}_i$ and an approximation of $\mathbfcal{T}_i$ calculated from the SVD of the mean adjacency matrix $\mathbfcal{T}_{(k)}$:
    \begin{equation}
        d_w = \sum_{k=1}^p \sum_{i: y_i \in C_k} ||\mathbfcal{T}_{i} - \mathbf{U}_{(k)}\mathbf{\Sigma}_i\mathbf{V}_{(k)}^{T}||^2_F
        \label{eq:my_WSS}
    \end{equation}
    where $||\cdot||_F$ is the Frobenius norm. The average matrix of the entire set of adjacency matrices, $\mathbfcal{T}_{(0)} = \frac1{n}\sum_{i=1}^n \mathbfcal{T}_{i}$, is decomposed $\mathbfcal{T}_{(0)} = \mathbf{U}_{(0)}\mathbf{\Sigma}_{(0)} \mathbf{V}_{(0)}^T.$ Then, the between-cluster distance metric is based on the difference between each $\mathbfcal{T}_{(k)}$ and an approximation of $\mathbfcal{T}_{(k)}$ calculated from SVD of the mean adjacency matrix for all sequences $\mathbfcal{T}_{(0)}$:
    \begin{equation}
        d_b = \sum_{k=1}^{p} n_k||\mathbfcal{T}_{(k)} - \mathbf{U}_{(0)}\mathbf{\Sigma}_{(k)}\mathbf{V}_{(0)}^T||^2_F
        \label{eq:my_BSS}
    \end{equation}
    We note that in equations \ref{eq:my_WSS} and \ref{eq:my_BSS} other matrix norms can be used instead of the Frobenius norm, which we choose due to its low computational cost. These two metrics can be combined in the same manner that the Calinski-Harabasz (CH) index \citep{CH1974} combines the sum of squares within cluster and sum of squares between cluster metrics,
    \begin{equation}
        D = \frac{d_b/(p -1)}{d_w/(n-p)}
        \label{eq:my_CH}
    \end{equation}
    This final metric can be maximized to determine tuning parameter values for clustering the set of sequences $\{y_i\}$, including $h$, the number of columns chosen from the SVD for clustering, as well as the optimal number of clusters.

 A direct matrix form translation of sum of squares within cluster and sum of squares between cluster is:
\begin{equation}
    WSS = \sum_{k=1}^p \sum_{i: y_i \in C_k} tr[(\mathbfcal{T}_i - \mathbfcal{T}_{(k)})^T(\mathbfcal{T}_i - \mathbfcal{T}_{(k)})] = \sum_{k=1}^p \sum_{i: y_i \in C_k} ||\mathbfcal{T}_i - \mathbfcal{T}_{(k)}||^2_F
    \label{eq:standard_wss}
\end{equation}
\begin{equation}
    BSS = \sum_{k=1}^{p} n_k tr[(\mathbfcal{T}_{(k)} - \mathbfcal{T}_{(0)})^T(\mathbfcal{T}_{(k)} - \mathbfcal{T}_{(0)})] = \sum_{k=1}^{p} n_k||\mathbfcal{T}_{(k)} - \mathbfcal{T}_{(0)}||^2_F
    \label{eq:standard_bss}
\end{equation}

This formulation of $WSS$ and $BSS$ could also be combined in the same manner as the CH index and used to evaluate clusters (and may be desirable to use depending on computational considerations). However, selecting the optimal number of clusters based on these formulations tends to select a higher number of clusters than Equation \ref{eq:my_CH}. In addition, rather than evaluating the numerical spread of matrices within and between clusters (Equations \ref{eq:standard_wss} and \ref{eq:standard_bss}), our proposed metric measures how well each cluster represents the sequences within it. In Equation \ref{eq:my_WSS}, $\mathbf{U}_{(k)}$ and $\mathbf{V}_{(k)}^T$ encapsulate the shared aspects of matrices within cluster $C_k$ while $\mathbf{\Sigma}_i$ represents the individual attributes of a single matrix, $\mathbfcal{T}_i$ \citep{Tang2009, Dong2012}. Then, if a sequence, $\mathbfcal{T}_i$, is well represented by a cluster, $C_k$, $\mathbfcal{T}_i \approx \mathbf{U}_{(k)}\mathbf{\Sigma}_i\mathbf{V}^{T}_{(k)}$ such that $||\mathbfcal{T}_i - \mathbf{U}_{(k)}\mathbf{\Sigma}_i\mathbf{V}_{(k)}^{T}||_F^2 \approx 0$.

\subsection{Computational Benefits}
Since all three methods implement a traditional clustering algorithm (e.g., hierarchical or k-means), the primary differences in computational requirements are due to the data preparation steps. For hierarchical clustering, the Levenshtein distance matrix must be calculated  which has complexity $O(n^2l^2)$ where $n$ is the number of sequences in the set and $l$ is the length of the sequences. There are two data preparation steps for nTreeClus, 1) a matrix segmentation process with complexity $n(l-k)$; and 2) the Random Forest algorithm with complexity $O(t\sqrt{l -k}n\log(n))$ where $k$ is the $k-$mer parameter and $t$ is the number of trees used within the Random Forest. The creators of nTreeClus recommend $k = \sqrt{n}$ and $t = 10$. Our method computes the SVD of the vectorized adjacency matrices which has complexity $O(\text{min}(nm^4, n^2m^2))$ where $m$ is the number of unique states in the set of sequences. In this application, most often $\text{min}(nm^4, n^2m^2) = nm^4$ and $m \ll l$. Thus, the complexity of our method is substantially less than the complexity of the distance matrix calculation. The complexity of nTreeClus and our method can be similar depending on the data attributes; however, the complexity of our method tends to be less than nTreeClus as sequence length increases for a given number of states $m$. In the analysis of Daynamica data in Section \ref{sec:data_app}, $m = 4$ for all sequences while $l = 288$ for day sequences and $l = 1440$ for week sequences; with these values the complexity of our method's data preparation step is less than that of nTreeClus.

\subsection{Weighting} \label{sect:weighting}

The adjacency matrix representation of a sequence gives equal weight to every state transition, regardless of the location in the sequence where the transition occurs or the states involved in the transition. However, it may be desirable to emphasize the transitions that occur within a given region of the sequence or transitions that involve a specific state. For example, a researcher may want to increase the relative influence of transitions that occur during standard commuting hours (6-9 a.m., 3-6 p.m.) to better distinguish individuals with different commuting patterns. 

For a set of sequences $\{y_i\}$, we can weight aspects of the adjacency matrix in order to increase their influence within the clustering procedure. Formally, for the transition from entry $j$ to entry $j+1$ in sequence $y_i$ we assign a corresponding weight $w_{j}$ such that the $(u,v)$ entry of the weighted adjacency matrix, $\mathbfcal{T}_i^w$, is constructed as $\mathbfcal{T}^w_{i_{u,v}} = \sum_{j = 1}^{l-1} w_j I[y_{i,j} = s_u, y_{i,j+1} = s_v]$. The weighted adjacency matrix is then scaled such that the entries in each each matrix still sum to $l -1$, $\mathbfcal{T}_i^w = \frac{\text{sum}(\mathbfcal{T}_i^w )}{l-1}\mathbfcal{T}_i^w.$ The rest of the clustering process remains the same, simply using the set of weighted adjacency matrices, $\{\mathbfcal{T}^w_i\}$. For the example previously mentioned, we can assign a weight $w^*$ to all transitions occurring between 6-9 a.m. and 3-6 p.m and a weight $w$ to transitions occurring at all other times such that $w^*/w > 1$ to emphasize the commuting hours. Note that the weights can also be changed continuously for each $j$ based on some criteria of interest. We illustrate an example of applying weights in Section \ref{sec:data_app}.

\section{Simulation}\label{sec:simulation}
\subsection{Methods}
Simulating sequences that are analogous to real data, in a manner that is parametrizable, is challenging and an active area of research \citep{huang2023}. Rather than attempt to replicate real data via simulation, we assess the performance of the clustering methods on sequences simulated from first, second, and fifth order Markov chains to explore the robustness of our method to deviations from the implicit first-order Markov assumption in the adjacency matrix representation. In addition, the transition probability matrices used to simulate the data are variations on the estimated transition probability matrices from Daynamica human activity data such that the simulated sequences have similarities to the empirical data.

To explore the performance of our method, sequences are generated from a mixture of data generating processes (i.e., true clusters), which differ according to the distribution of states present in the sequences and the probability of state transitions (or equivalently, the contiguous average state duration) as these are the primary attributes we want to distinguish in the human activity data. In all cases, data is generated from an equal mix of the true clusters. For clusters determined by differences in the states present, we create three scenarios of increasing difficulty based on the overlap of states present between clusters (with four possible states: $A, \: B,\: C,$ and $D$): 1) Low overlap: two clusters with no overlap in states; 2) Medium overlap: three clusters with 1 pairwise state overlapping; 3) High overlap: four clusters with 2 pairwise states overlapping. For clusters determined by differences in average contiguous state duration, we create three scenarios of increasing difficulty based on the distance between the cluster generating transition probability matrices (larger distances between the transition probability matrices correspond to lower overlap clusters). All clusters have three states ($A,\:B,$ and $C$) and we explore varying the average contiguous duration of one state, two states, and three states by cluster for each overlap scenario. Supplementary Figures 1-3 provide examples of the sequences within each cluster. For all simulation scenarios, we simulate 500 datasets with 250 sequences of length 500; these values are chosen in part to ensure reasonable computation time for comparison methods.

We compare our method, adjacency matrix decomposition clustering, to agglomerative hierarchical clustering using the average linkage function of the Levenshtein sequence distance matrix and to nTreeClus by \cite{jahanshahi_ntreeclus_2022}. For hierarchical clustering, we select the optimal number of clusters by the Dunn Index \citep{Dunn1974}. We run the clustering algorithms (k-means and hierarchical) ten times on each simulated dataset and select the best set of clusters through the respective cluster evaluation metric. For nTreeClus, we use the implementation provided by \cite{jahanshahi_ntreeclus_2022} with default parameters; the implementation provides clustering results from four method variations and we select the optimal clustering (method variation and cluster number) by the Average Silhouette Width as recommended. For each simulated dataset, we calculate the percent accuracy of the methods when selecting the true number of clusters by taking the highest accuracy possible given the constraint that the calculated clusters must have a bijective relationship with the true clusters. In addition, we determine the optimal number of clusters by the respective criteria for each method. For each simulation scenario, we then assess the clustering methods by the average percent accuracy and the mode optimal number of clusters in comparison to the true number of clusters across simulated datasets.

\subsection{Results}
Across all scenarios where sequences are simulated from first order Markov-chains, adjacency matrix decomposition has similar or higher clustering accuracy than hierarchical clustering or nTreeClus (Table \ref{table:sim_first_res}). Performance for all methods is lowest for scenarios where the average duration of only one state varies by cluster, indicating a potential limit to the type of sequence differences these methods can distinguish. Our method shows the largest benefit over hierarchical clustering and nTreeClus in high overlap scenarios where clusters differ by the average state duration of two or three states; the accuracy of adjacency matrix decomposition clustering is 0.15 - 0.18 higher in these scenarios. Therefore, as the clusters become less distinct, our method significantly outperforms competing methods across the scenarios. In selecting the optimal number of clusters, all methods tend to perform similarly. In all but one case, all methods select either the correct number of clusters or are off by one cluster. The methods have similar performance across all scenarios when the sequences are simulated from second-order Markov Chains (Supplementary Table 1).

\begin{table}[h!]
\centering
\small
\caption{ \small Simulation results for sequences simulated by first-order Markov chains.}
\begin{tabular}{lllll|llll}
\hline
&& \multicolumn{3}{c}{\% Accuracy (SD)} & \multicolumn{4}{c}{No. Clusters} \\
Overlap & Type & 
 Hier.&  AMDC & nTreeClus 
  & True & 
  Hier.&  AMDC & nTreeClus \\
\hline
Low & State & 1.00 (0.00) & 1.00 (0.00) & 1.00 (0.00) & 2 & 2 & 2 & 2\\

Medium & &  0.73 (0.07) & 0.83 (0.03) & 0.79 (0.04) & 3 & 2 & 6 & 2\\

High & & 0.63 (0.06) & 0.66 (0.06) & 0.61 (0.04) & 4 & 2 & 2 & 2\\

Low & Dur. 1& 0.63 (0.04) & 0.71 (0.06) & 0.69 (0.05) & 3 & 2 & 2 & 2\\

Medium &  & 0.61 (0.04) & 0.69 (0.04) & 0.68 (0.04) & 3 & 2 & 2 & 2\\

High & & 0.51 (0.07) & 0.58 (0.03) & 0.57 (0.03) & 3 & 2 & 2 & 2\\

Low & Dur. 2 & 0.69 (0.02) & 0.98 (0.01) & 0.78 (0.06) & 3 & 2 & 2 & 2\\

Medium & & 0.68 (0.02) & 0.95 (0.05) & 0.71 (0.04) & 3 & 2 & 2 & 2\\

High & & 0.59 (0.06) & 0.77 (0.04) & 0.60 (0.06) & 3 & 2 & 2 & 2\\

Low & Dur. 3&1.00 (0.01) & 1.00 (0.00) & 0.94 (0.03) & 3 & 3 & 4 & 3\\

Medium & & 0.95 (0.05) & 0.98 (0.03) & 0.83 (0.05) & 3 & 2 & 2 & 2\\

High & & 0.65 (0.04) & 0.82 (0.04) & 0.67 (0.04) & 3 & 2 & 2 & 2\\
\hline
\end{tabular}
\vspace{-0.1in}
\begin{flushleft}
\small Note: ``Overlap'' indicates the extent of the overlap in sequences between the true clusters (high overlap corresponds to less distinct clusters), ``Type'' indicates whether the true clusters differ by the states present or the average state duration.
\end{flushleft}
\label{table:sim_first_res}
\end{table}

When sequences are simulated from fifth-order Markov chains, our method still has similar or superior performance to hierarchical clustering and nTreeClus for all scenarios except one (Table \ref{table:sim_fifth_res}). Across most scenarios, our method's accuracy is within 0.01-0.02 of the accuracy from the first-order Markov chain simulated sequences; in two scenarios our method's accuracy is higher for fifth-order Markov chain simualted sequences than for first-order. In the medium and high overlap scenarios where three state durations vary by cluster, the accuracy of our method decreases by 0.04 and 0.06 respectively. Despite this, adjacency matrix decomposition clustering maintains at least 0.08 higher accuracy than hierarchical clustering and nTreeClus. In nearly all simulation scenarios the same number of optimal clusters is selected as for the first-order Markov chain simulated sequences. The performance of adjacency matrix decomposition clustering compared to competing methods as well as the minimal decreases in accuracy for fifth-order Markov chain simulated sequences demonstrates that our method is robust to deviations from the first-order Markov assumption.

\begin{table}[h!]
\centering
\small
\caption{ \small Simulation results for sequences simulated by fifth-order Markov chains.}
\begin{tabular}{lllll|llll}
\hline
&& \multicolumn{3}{c}{\% Accuracy (SD)} & \multicolumn{4}{c}{No. Clusters} \\
Overlap & Type & 
 Hier.&  AMDC & nTreeClus 
  & True & 
  Hier.&  AMDC & nTreeClus \\
\hline
Low & State& 1.00 (0.00) & 1.00 (0.00) & 1.00 (0.00) & 2 & 2 & 2 & 2\\

Medium & &0.73 (0.08) & 0.82 (0.04) & 0.79 (0.05) & 3 & 2 & 6 & 2\\

High & & 0.62 (0.06) & 0.65 (0.09) & 0.62 (0.06) & 4 & 2 & 2 & 2\\

Low & Dur. 1 &0.62 (0.04) & 0.70 (0.05) & 0.67 (0.05) & 3 & 2 & 2 & 2\\

Medium & & 0.60 (0.06) & 0.67 (0.04) & 0.66 (0.04) & 3 & 2 & 2 & 2\\

High & & 0.50 (0.09) & 0.57 (0.03) & 0.57 (0.03) & 3 & 2 & 2 & 2\\

Low & Dur. 2 &0.69 (0.02) & 0.98 (0.02) & 0.99 (0.01) & 3 & 2 & 2 & 3\\

Medium & & 0.68 (0.02) & 0.96 (0.03) & 0.87 (0.12) & 3 & 2 & 2 & 2\\

High & & 0.58 (0.08) & 0.78 (0.04) & 0.63 (0.05) & 3 & 2 & 2 & 2\\

Low & Dur. 3 & 0.93 (0.09) & 0.98 (0.01) & 0.96 (0.03) & 3 & 2 & 2 & 3\\

Medium & & 0.74 (0.07) & 0.94 (0.02) & 0.85 (0.05) & 3 & 2 & 2 & 2\\

High & & 0.63 (0.04) & 0.76 (0.04) & 0.68 (0.04) & 3 & 2 & 2 & 2\\
\hline
\end{tabular}
\vspace{-0.1in}
\begin{flushleft}
\small Note: ``Overlap'' indicates the extent of the overlap in sequences between the true clusters (high overlap corresponds to less distinct clusters), ``Type'' indicates whether the true clusters differ by the states present or the average state duration.
\end{flushleft}
\label{table:sim_fifth_res}
\end{table}

\section{Sequential Human Activity Data}\label{sec:data_app}

Our motivating data come from the COVID-19 Implications on Public Transportation Study conducted between March 2021 and June 2021 in the Minneapolis-St. Paul metropolitan area \citep{fan2022}. The study recruited 339 participants through various digital marketing tools, such as social media and website advertising, in order to remotely recruit a diverse sample of individuals. The study consisted of two components: an intake survey and the use of a smartphone app to collect data on daily activities. The intake survey was administered virtually and collected demographic, health, and transportation information about the participants. The smartphone app, called Daynamica (previously SmarTrAC \citep{Fan2015}), used mobile sensing to automatically detect activities and trips in real time and collected survey data on subjective trip/day information (e.g., trip companionship, general trip experience, how COVID-19 influenced plans, etc.) for at least 14 days. Of the recruited participants, 154 completed the full two weeks of data collection. Our interest is in identifying common time use patterns (clusters) from this smartphone-derived daily activity data and associating these time use patterns with reported COVID-19 concerns.

The 24-hour data from the multi-day collection period is aggregated into five-minute intervals, creating sequences with 288 positions. Each position contains the state where the individual spent the majority of the corresponding five-minute interval. Daynamica categorizes the data into 14 activity states (such as work, home, travel by bus, bike, car, eating out, shopping, etc.). However, we use the following possible states: H - Home, W - Work, T - Transport, and O - any other activity (Out). We chose to use this aggregation of states to facilitate a more meaningful interpretation of the time use patterns. Any sequences with missing data are removed from the analysis. Most individuals have fewer than 20 day sequences within the data (median: 13 sequences); however, to ensure that certain individuals are not over-represented in the data we randomly select at most 20 sequences per individual. Days with over 90\% of time spent at work, traveling, or out are also removed from the analysis as they likely contain data collection errors.

The final dataset of 24-hour sequences contains 2,229 sequences from 187 individuals. While sequences from the same individuals are likely correlated, we are interested in identifying common patterns in the daily sequences themselves, not common patterns among individuals. However, having multiple day sequences per individual motivates identifying sequence patterns for longer time periods, such as a week. Individuals with complete, consecutive data for Monday-Friday are combined to create week sequences, where each week sequence has length 1440. There are 126 five-day sequences from 102 individuals. We perform hierarchical clustering, nTreeClus, and the proposed clustering method on all day and week sequences. We explore the stability of the day sequence clusters from our method through a bootstrapping stability procedure based on the Jaccard index \citep{yu_bootstrapping_2019}. Further details of the stability analysis are in Supplementary Section 1. We also demonstrate an example of weighted adjacency matrix decomposition clustering on the day sequences. We explore three different relative weights and present the clustering results for the intermediate weight. We use relative weights of 1.5, 2, and 2.5 for the 9AM-5PM time period within the sequences. Note that these weights are chosen for illustrative purposes; selection of weights for a given research question would require additional considerations.

In addition to the clusters themselves, we want to understand in how the day patterns identified by our method are related to concerns about contracting COVID-19. In the end of day surveys given within the Daynamica app, we are interested in one question: ``Indicate how much you agree with the following statement: Overall, I am concerned with having contracted Coronavirus today". We recode this question as a binary response; all levels of disagreement are encoded as disagree and all other levels are encoded as agree. We then fit a logistic mixed-effect models for this binary response with cluster assignment as a categorical predictor and a random intercept for each individual. From these models we can determine how daily activity patterns are related to COVID-19 concern.

\subsection{Day Sequence Analysis}
\subsubsection{Clustering Results}
\begin{figure}
    \centering
    \includegraphics[scale = .15]{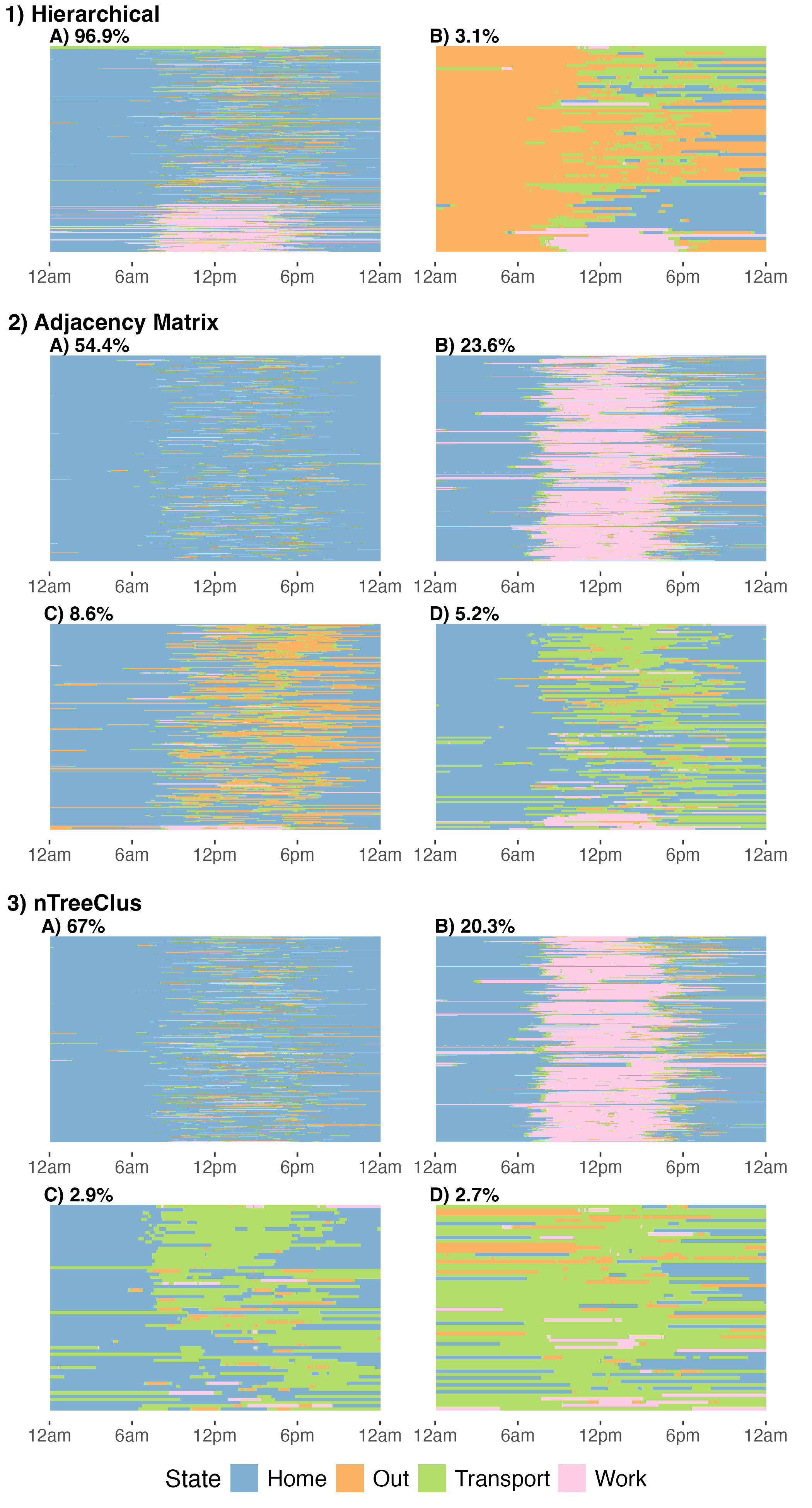}
    \caption{\footnotesize Day sequence clusters by clustering method. Panel one depicts the two optimal clusters from hierarchical clustering, panel two and three depict the four largest of the eight optimal clusters from adjacency matrix decomposition clustering and nTreeClus respectively. Labels indicate the percent of all sequences in the respective cluster. Within each sub-figure, each row depicts a single sequence; row height has been standardized such that the sub-figure is the same height for all clusters.}
    \label{fig:day_clusts}
\end{figure}

Hierarchical clustering of all day sequences results in two optimal clusters, as chosen by the Dunn Index (Figure \ref{fig:day_clusts}, Panel 1). Almost all sequences (96.9\%) are in cluster A, while cluster B contains sequences with a majority of Out states. In contrast, both adjacency matrix decomposition clustering and nTreeClus select eight optimal clusters, where the four largest clusters are shown in Figure \ref{fig:day_clusts}, Panels 2 and 3. For both methods the two largest clusters contain similar sequences; cluster A contains sequences have predominantly Home states while cluster B contains sequences with mostly Work and Home states. With nTreeClus, almost 90\% of sequences are clustered within these two clusters, all other clusters have few sequences. Clusters C and D comprise of sequences that have a many Transport states; around 30-50\% of each sequence is Transport states in cluster C, while almost the entire sequence is Transport states in cluster D. The remaining nTreeClus clusters distinguish similar patterns in both the Work and Out states (Supplementary Figure 4). With adjacency matrix decomposition clustering, the sequences in clusters C and D are similar to cluster A but have more Out and Tranport states. The other four remaining clusters contain sequences with mostly Work, Transport, or Out states (Supplementary Figure 5).

While hierarchical clustering groups most sequences into a single cluster, adjacency matrix decomposition clustering and nTreeClus both separate those sequences into multiple meaningful clusters. For example, clusters A and B separate day patterns in which people spend most of their days at home and day patterns in which people go to work full-time, an important distinction, while hierarchical clustering is unable to separate these two day types. Cluster D from our method and cluster C from nTreeClus are similar; however, while nTreeClus distinguishes sequences that either have very few transport states (cluster A) or sequences that have $\frac13$ to $\frac12$ transport states (cluster C), our method can distinguish sequences with smaller differences in the number of transport states as seen in cluster D. This is consistent with the simulation results from Section \ref{sec:simulation} where our method considerably outperformed nTreeClus in high cluster overlap scenarios. There is no analog to cluster C from our method in the nTreeClus clustering results. With our method, clusters C and D distinguish daily time use patterns in which people spent significant time moving from place to place from other daily time use patterns in which people were able to limit their travel time and yet spent significant time outside home and work. The clusters from adjacency matrix decomposition clustering are relatively stable with a mean and median stability of $0.79$ and $0.88$ across all observations. Additional cluster-level stability analysis results are presented in Supplementary Section 1.  

Of note, even when eight clusters are selected in hierarchical clustering, one cluster still contains a majority of sequences (92.6\%) and workdays are not separated from other day types (Supplementary Figure 6). When implementing nTreeClus, the two best method variations have similar performance as measured by the Average Silhouette Width (0.795 and 0.792); one method incorporates the positioning of the states within the sequence and the other does not. The cluster results from the method that does not incorporate position have already been discussed (Figure \ref{fig:day_clusts} and Supplementary Figure 4). The method that includes state positioning performs poorly; only two optimal clusters are selected and one cluster contains a majority (92\%) of sequences (Supplementary Figure 7). These two clustering results are incredibly different despite the similar Average Silhouette Width which indicates a potential additional limitation of the nTreeClus method and implementation for human activity data.

\subsubsection{Relation to COVID-19 Concerns} 
In the mixed effects model, almost all day patterns are significantly associated with a higher odds of being concerned about contracting COVID-19 compared to days spent mostly at home (Supplementary Table 2). For clusters C, E, and F (Supplementary Figure 5), the odds of being concerned about contracting COVID-19 are nearly four times higher than for those in cluster A. Clusters C, E, and F contain sequences with most time spent in Home and Out states, Transport states, and Out states respectively. Of note, clusters C and F have similar odds of COVID-19 concern despite the days in cluster F having substantially more time spent in Out States; this may indicate that individuals' concern with contracting COVID-19 does not increase linearly with time spent outside of home, work, and transport. Instead, once an individual spends some time outside of home, work, and transport they may have similar concerns about contracting COVID-19 as if they spent nearly the whole day out. In contrast, the day pattern with nearly all time spent in Transport states (cluster E) does have a higher odds of COVID-19 concern compared to the day patterns with only some time spent in Transport states (cluster D). Day patterns that include substantial time spent at work (clusters B and G) have the smallest increase in odds of being concerned about contracting COVID-19 compared to days spent at home. These results indicate that individuals are less concerned about contracting COVID-19 when at work compared to transport and other activities. In addition, individuals have a higher odds of being concerned about contracting COVID-19 when they spend moderate time in Out states (cluster C) compared to when they spent moderate time in Transport states (cluster D).

\subsection{Week Sequence Analysis}
Hierarchical clustering of week sequences results in four optimal clusters; however, three of the clusters contain only one sequence such that one cluster contains almost all (97.6\%) sequences. nTreeClus also performs poorly; the largest of the three optimal clusters (77\%) contains both sequences with mostly Home states and sequences with a considerable number of Work states (Supplementary Figure 8). Adjacency matrix decomposition clustering selects six optimal clusters where the sequences in each cluster differ primarily by the number of Home and Work states (Supplementary Figure 9). We can calculate the contribution of each adjacency matrix entry to the column vectors of $\mathbf{V}$ to explicitly understand what factors are influencing the clusters. Formally, each row of $\mathbfcal{M}_c$ corresponds to an entry of the adjacency matrix, and we can identify the percent contribution of each of row to the columns of $\mathbf{V}$ with $\mathbf{C} = (\mathbfcal{M}_c\mathbf{V})^{\circ 2}(100/(\mathbf{\Sigma}^2)$ where $\circ$ indicates element-wise operation.

\begin{figure}[h!]
    \centering
    \includegraphics[scale = .2]{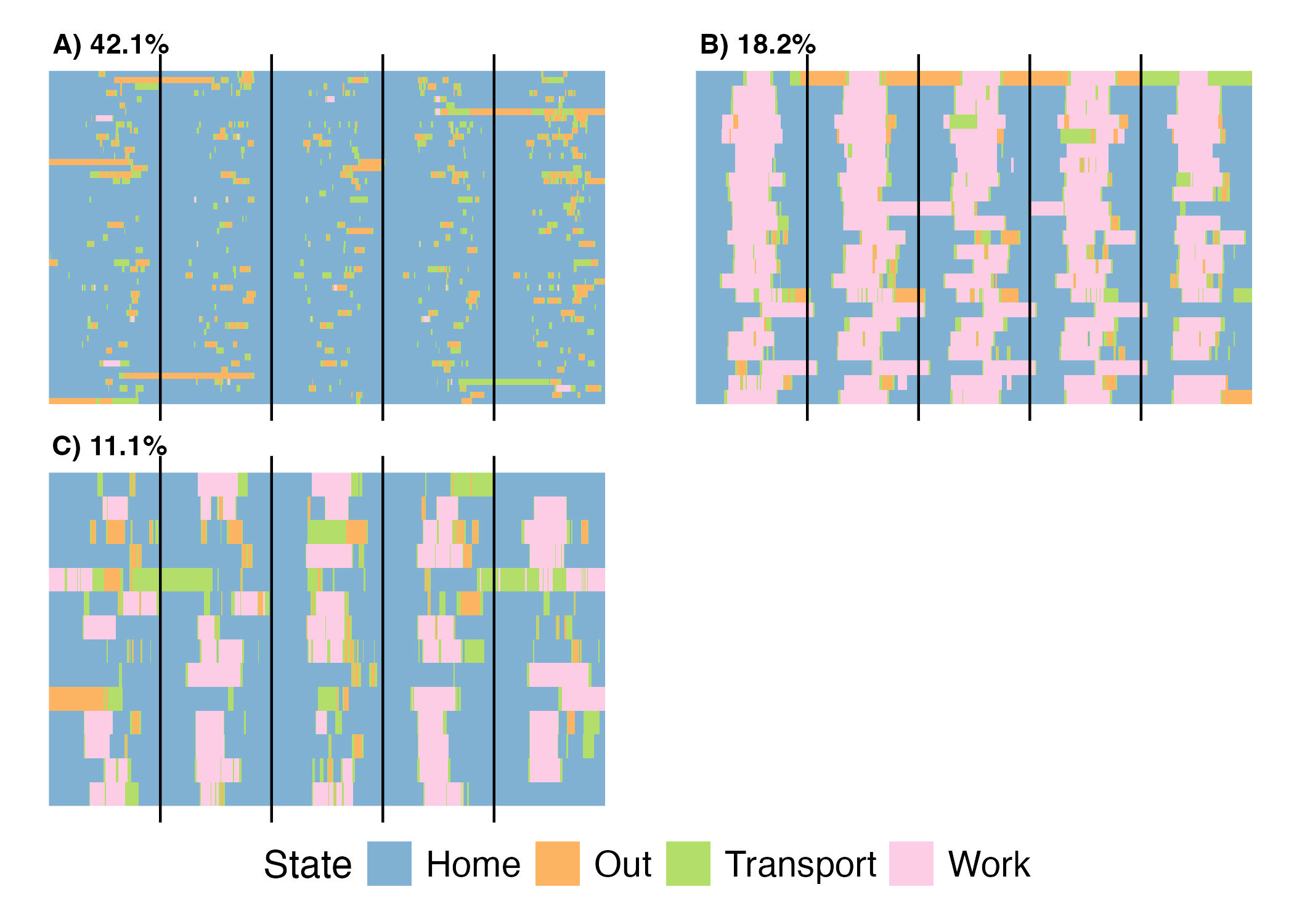}
    \caption{\footnotesize Monday-Friday sequence clusters from adjacency matrix decomposition clustering. Black vertical lines separate each of the five days. This figure presents three of the five optimal clusters which predominantly distinguish days by amount of time spent at home and work. Labels indicate the percent of all sequences in the respective cluster. Within each sub-figure, each row depicts a single sequence; row height has been standardized such that the sub-figure is the same height for all clusters.}
    \label{fig:week_clusts}
\end{figure}

In this case, the optimal clustering uses the first two column vectors of $\mathbf{V}$; the Home duration entry of the adjacency matrix contributes 91.5\% to the first vector and the Work duration entry contributes 91.1\% to the second vector, which confirms the primary influence of the Home and Work states on the clustering. Figure \ref{fig:week_clusts} shows three of the clusters: cluster A contains weeks that were spent mostly at home, cluster B contains weeks of five in-person, eight-hour workdays, and cluster C contains week patterns where either individuals worked five shorter work days or individuals worked fewer than five in-person, eight-hour work days. Essentially, each cluster contains week patterns of people who stay at home (A), work in-person full-time (B), and work part-time or hybrid (C). With these longer sequences, hierarchical clustering and nTreeClus are unable to detect and separate any substantial differences in the sequences. However, our method results in clusters with clear interpretations in the context of Monday-Friday time use. 

\subsection{Weighted Day Sequence Analysis}
\label{sec:weighted_day_seq_res}
\begin{figure}[h!]
    \centering
    \includegraphics[scale = .2]{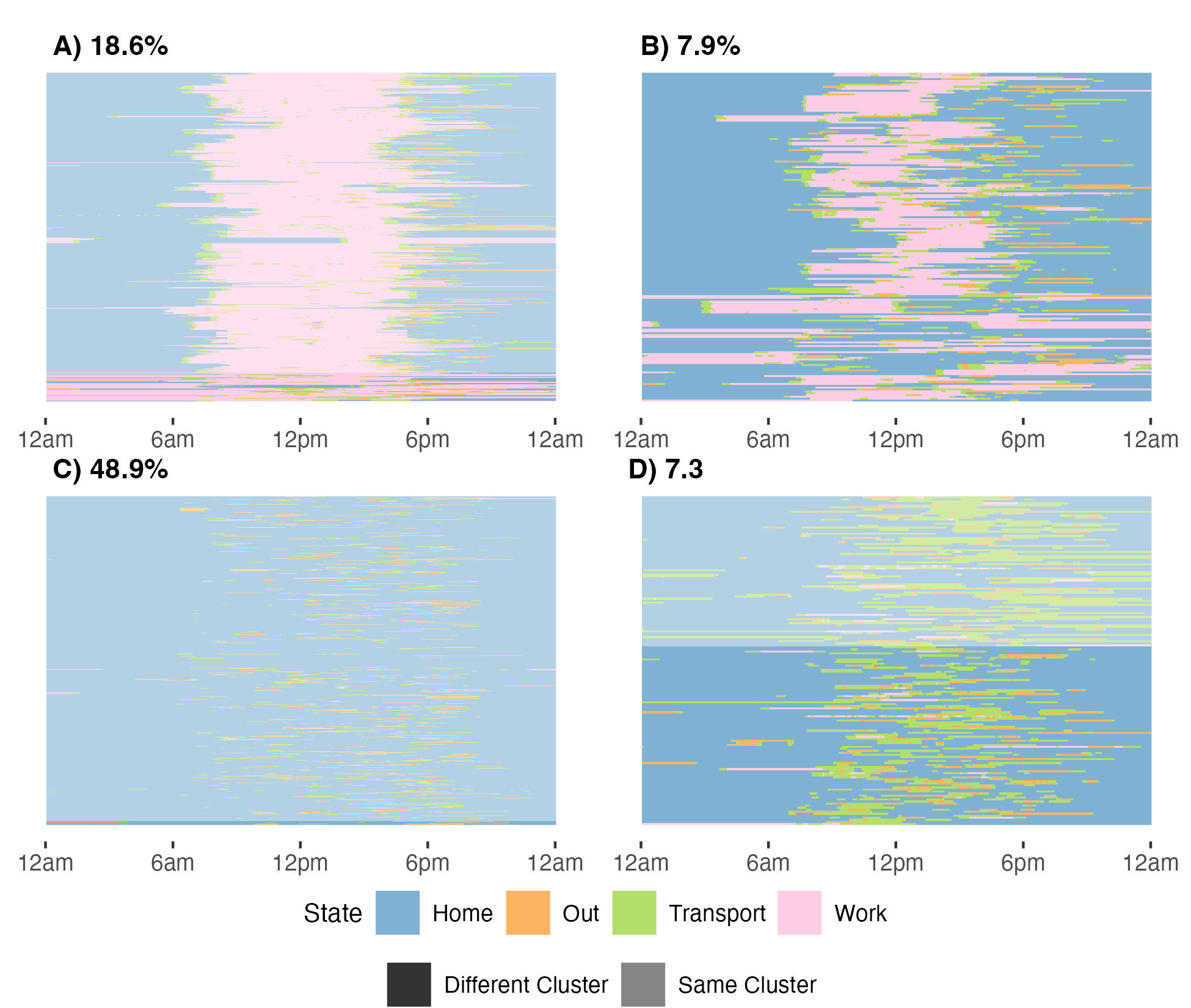}
    \caption{\footnotesize 9AM-5PM weighted day sequence clusters from adjacency matrix decomposition clustering. Darker shading distinguishes the sequences that are in a different cluster than in the original day sequence set of clusters in Figure \ref{fig:day_clusts} and supplementary Figure 3. This figure presents four out of the eight optimal clusters. Labels indicate the percent of all sequences in the respective cluster. Within each sub-figure, each row depicts a single sequence; row height has been standardized such that the sub-figure is the same height for all clusters.}
    \label{fig:9_5day}
\end{figure}

When weighting the 9AM-5PM portion of each sequence by two, we still obtain eight optimal clusters from adjacency matrix decomposition clustering (Supplementary Figure 10). Figure \ref{fig:9_5day} demonstrates how some of the clusters change from those in Figure \ref{fig:day_clusts}. One cluster that does not have a parallel from the original day clusters is cluster B in Figure \ref{fig:9_5day}. This cluster  contains days where individuals worked eight hours outside of the 9AM-5PM time or worked fewer than eight hours; this is in comparison to cluster A which contains days where people worked in the 9AM-5PM time frame. The adjacency matrix representation retains little information about the ordering of states, and therefore the original clustering places all eight-hour workday sequences in the same cluster regardless of the time the work occurs. With up-weighting 9AM-5PM, the clustering method is able to distinguish days with abnormal work times. In addition, a portion of sequences originally in the Home sequence cluster (cluster A in Figure \ref{fig:day_clusts}) with some Transport states in the 9AM-5PM time frame are now in the Home and Transport sequence cluster (cluster D in Figure \ref{fig:9_5day}) after weighting. When using relative weights of 1.5 and 2.5 we see similar results, except that for a weight of 1.5 work days are split predominantly by the amount of time spent working rather than time of day the work occurred (Supplementary Figures 11 and 12). While the weighted clustering maintains many of the original clusters, it also identifies a new cluster of abnormal work days and creates subtle changes in the original clusters specific to the weighted time period. 

\subsection{Computational Benefits}
Adjacency matrix decomposition clustering data preparation (calculation of adjacency matrices and SVD) is almost 2000 times faster than calculating the day sequence distance matrix and 1.85 times faster than the matrix segmentation process in nTreeClus (Supplementary Figure 13). For the entire data preparation and clustering procedure for day sequences, our method is computationally 148.4 and 10.2 times faster than hierarchical clustering and nTreeClus respectively. Our method has similar computational benefits over hierarchical clustering and nTreeClus in the week sequence analysis (1305.9 and 12.9 times faster for data preparation, 95.5 and 22.3 times for the entire procedure). The computational improvements with respect to nTreeClus are larger for the week sequences due to the fact that these sequences are substantially longer than the day sequences. Regardless, performing the proposed method is 10 to 145 times faster than performing hierarchical clustering or nTreeClus on this data; adjacency matrix decomposition clustering provides considerable computational improvements over both methods, especially for long sequences.

\section{Discussion}\label{sec:discussion}
There are few methods for clustering sequential human activity data that 1) leverage all the information the data provides and 2) are computationally efficient for large datasets. We improve on standard distance- and feature-based sequence clustering methods by eliminating sequence length as a computationally limiting factor. Our method represents each sequence as an adjacency matrix and clusters the sequences through a decomposition of these adjacency matrices. We also propose a new cluster evaluation metric that leverages the adjacency matrix structure. The ability to weight sequence sections of interest also allows researchers to explore specific questions of interest while still clustering on the entire, rather than a subset, of the sequence. 

In addition to lowering the computational cost, adjacency matrix decomposition clustering generates a more complete and nuanced understanding of different activity or time use patterns than both hierarchical clustering and nTreeClus. When associating day patterns identified by our method with concern for contracting COVID-19, we find that most day patterns are associated with an increased odds of COVID-19 concern in comparison to days spent at home. These results could inform targeted public health messaging to address the discrepancies between individuals' perceived and real risk of contracting COVID-19 in their daily activities. In addition, perceived risk is a key factor motivating protective behavioral changes \citep{brewer_risk_2004}; thus, these results may indicate that individuals are more likely to reduce even a moderate amount of activities outside of work or transport due to the higher perceived concern. 

The proposed method uses SVD, a Euclidean space technique, on compositional data (a simplex space) since each column in $\mathbfcal{M}_c$ sums to the same value and is constrained by the length of the sequence. While there are methods to transform compositional data into the Euclidean space, most methods (e.g., log-ratio transform) rely on the assumption that there are not true zeros within the data \citep{Greenacre2021}. In this data, some sequences never contain a particular state transition which yields true zeros in the adjacency matrix. In addition, we found that using compositional data techniques with a true zero correction factor in this stage led to poorer final clustering performance compared to standard SVD.

The performance of adjacency matrix decomposition clustering relies on the fact that the adjacency matrix retains most of the relevant information about the full sequence. Consistent with other feature-based methods, we are essentially proposing the adjacency matrix as the feature of interest that drives the clustering of sequences. In addition, our method has similarities to graph clustering methods. The most successful graph clustering methods identify patterns or substructures within a set of graphs and then cluster the graphs based on the similarities of these substructures \citep{aggarwal_survey_2010, piernik_clustering_2016}. These methods mine frequent structural patterns within a dataset, however the specific structural measures of interest are often defined a priori as with most feature-based clustering methods. We also note that the adjacency matrix proposed in Section \ref{sec:defining_adj_mat} can be thought of as a graph; thus, graph clustering methods could be used here though one would need to define the relevant substructures for this application. 

One limitation of all feature-based methods is that information is lost when distilling sequences or graphs into a smaller set of features. There are three main areas where the adjacency matrix representation may fail to retain important information about a sequence. Since the adjacency matrix implicitly makes a first-order Markov assumption, this representation could not be sufficient for sequences whose true process is a higher order Markov chain. However, we validated the robustness of our method to higher-order Markov chains in Section \ref{sec:simulation}. The adjacency matrix also retains little information about the ordering of states within the sequence as the matrix essentially collapses the temporal component of the data. While this indicates that our method may have difficulties distinguishing sequences that primarily differ in the order of states, the proposed weighting framework could be used to place more emphasis on certain sections within the data to differentiate state order as demonstrated in Section \ref{sec:weighted_day_seq_res}. Even so, we acknowledge that more work is needed to determine the process for choosing specific weights and assessing their impact on the clustering results. Finally, our method can have difficulties distinguishing sequences where multiple states' contiguous duration differs in a similar way (e.g.,  $AAAAAAAAABBBBBBBBB$ and $AAAABBBBAAAAABBBBB$) as there are only slight changes in the adjacency matrix between these types of sequences. However, sequences do not often differ this way in human activity data and therefore differentiating these sequences is of little interest.

\section*{Funding}
This material is based upon work supported by the National Science Foundation Graduate Research Fellowship Program under Grant No. 2237827.

\section*{Disclosure Statement}
Drs. Julian Wolfson and Yingling Fan are co-founders of Daynamica, Inc. They and the University of Minnesota hold equity and royalty interests in Daynamica pursuant to the exclusive license agreement involving the software program that provides smartphone solutions for travel and activity capturing. These interests have been reviewed and managed by the University of Minnesota in accordance with its Conflict of Interest policies.

\bibliographystyle{apalike} 
\bibliography{refs}

\pagebreak
\setcounter{figure}{0}
\setcounter{table}{0}

\section*{Supplementary Materials}

\begin{figure}[bp]
    \centering
    \subfloat[States present vary ]{\includegraphics[scale = .8]{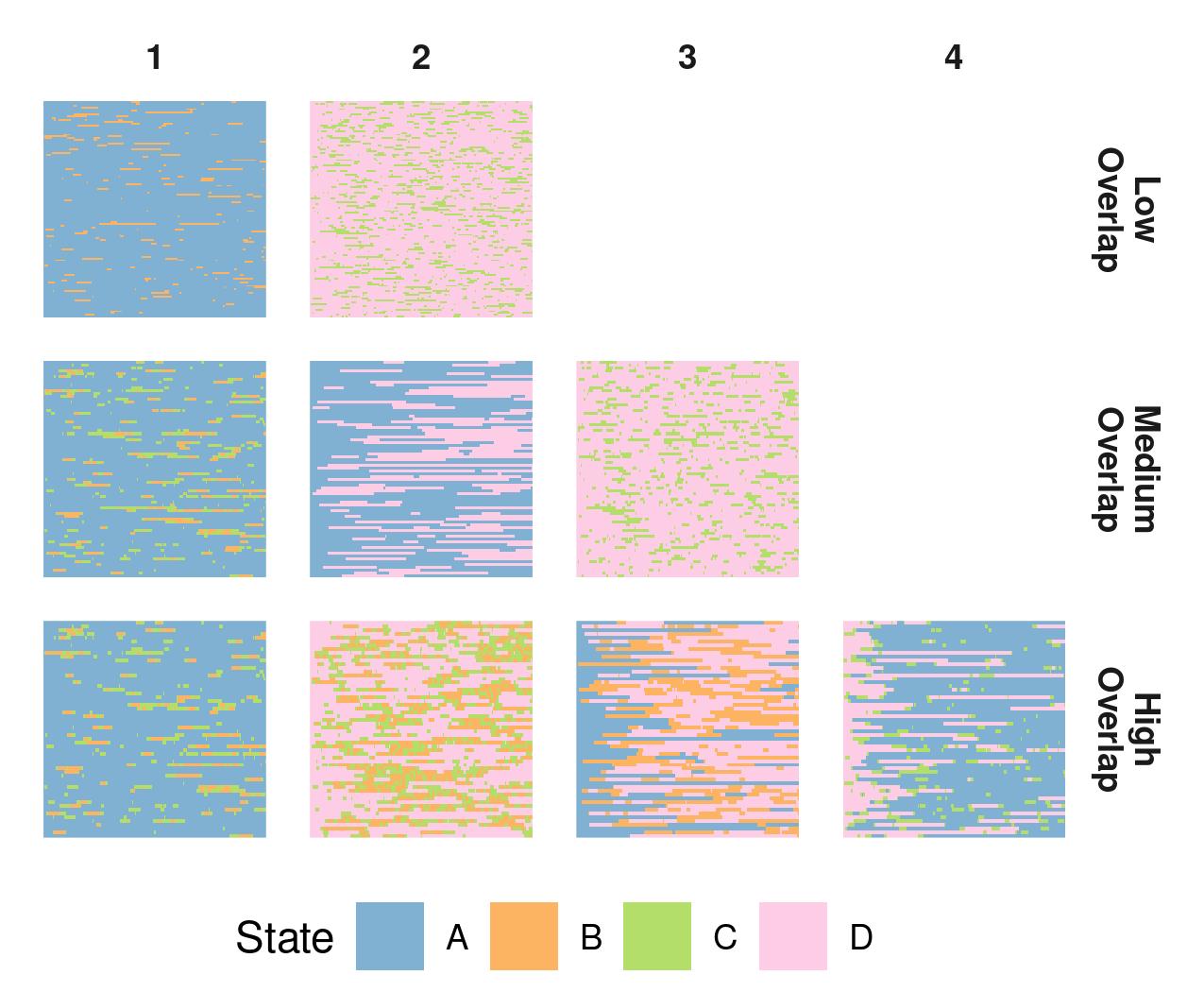}} 
    \subfloat[1 state duration varies]{\includegraphics[scale = .8]{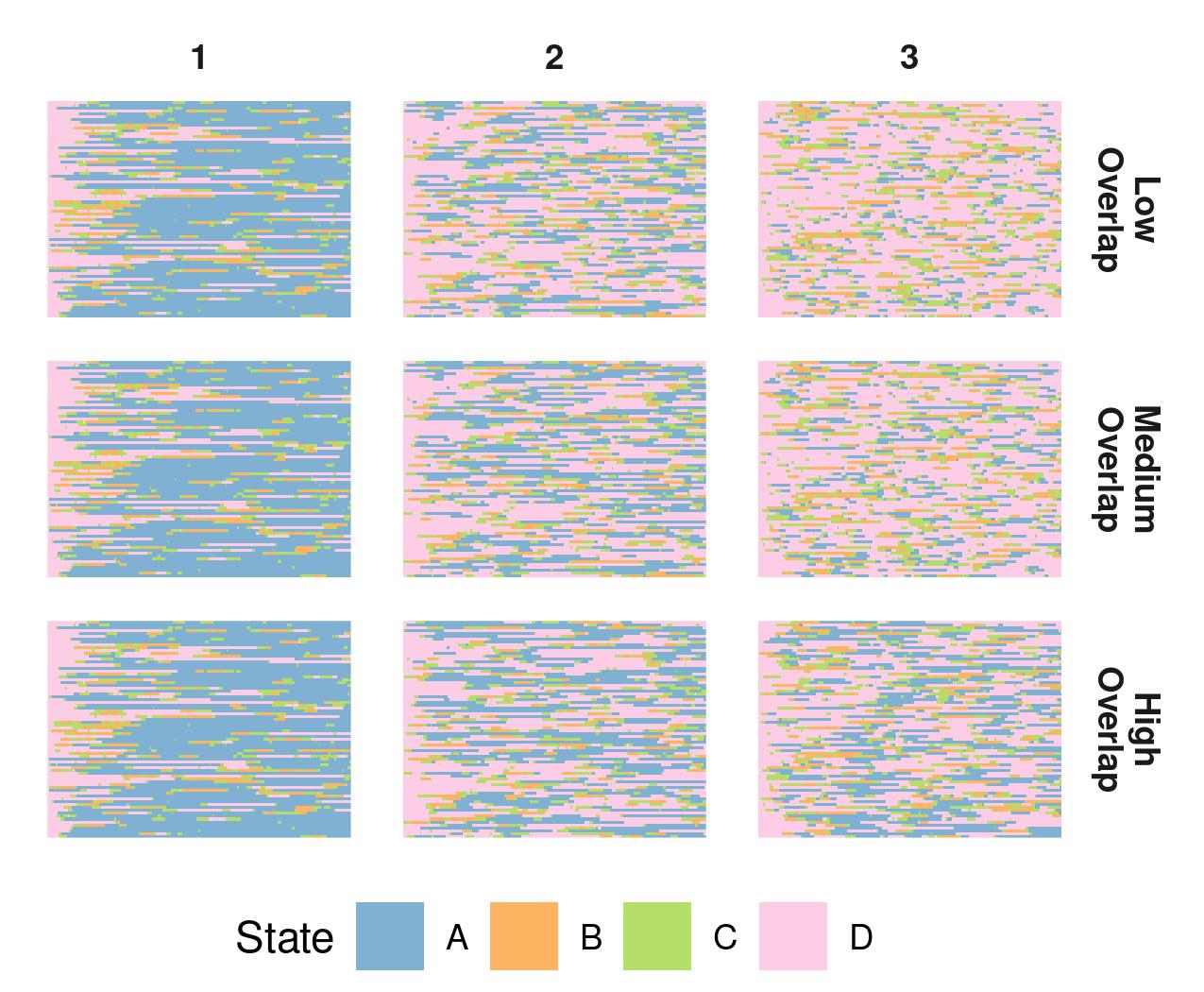}} \\
    \subfloat[2 state durations vary]{\includegraphics[scale = .8]{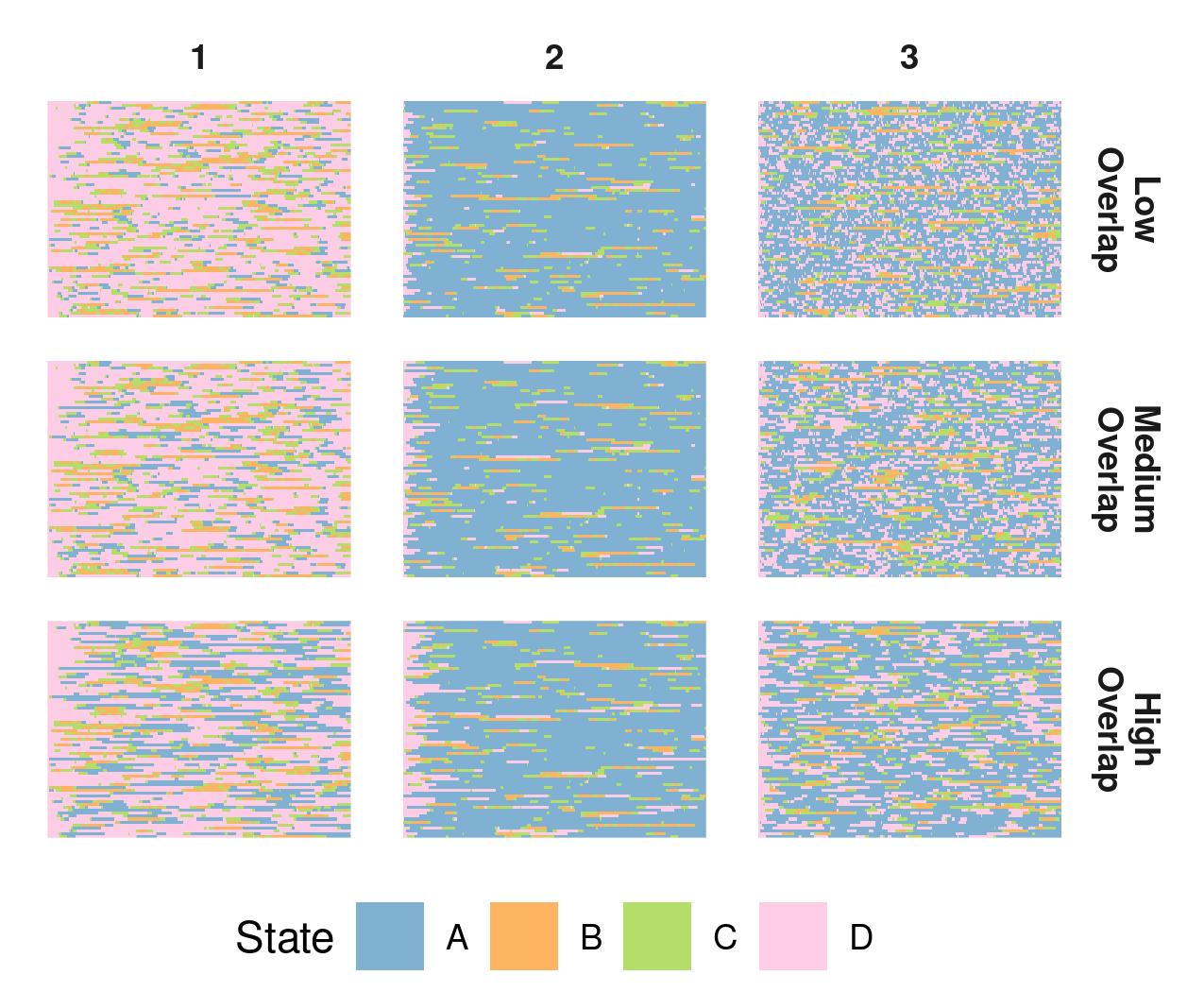}} 
    \subfloat[3 state durations vary]{\includegraphics[scale = .8]{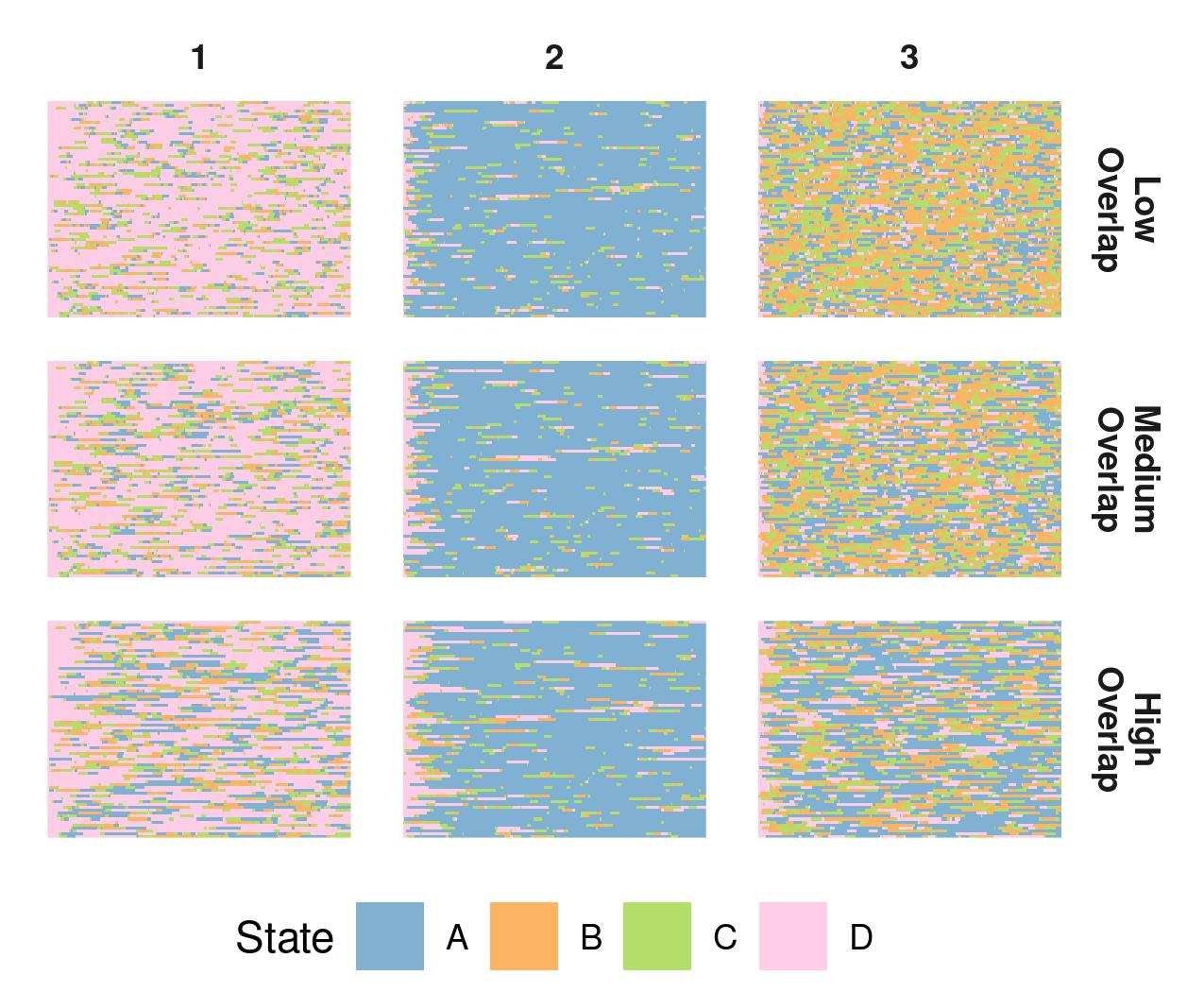}} 
    \caption{\small One sample set of simulated sequences from first-order Markov Chains (see Table 1 for corresponding results). Columns differentiate each cluster and rows differentiate simulation scenarios by overlap level.}
    \label{fig:sup_sim_order1}
\end{figure}

\begin{figure}[bp]
    \centering
    \subfloat[States present vary ]{\includegraphics[scale = .8]{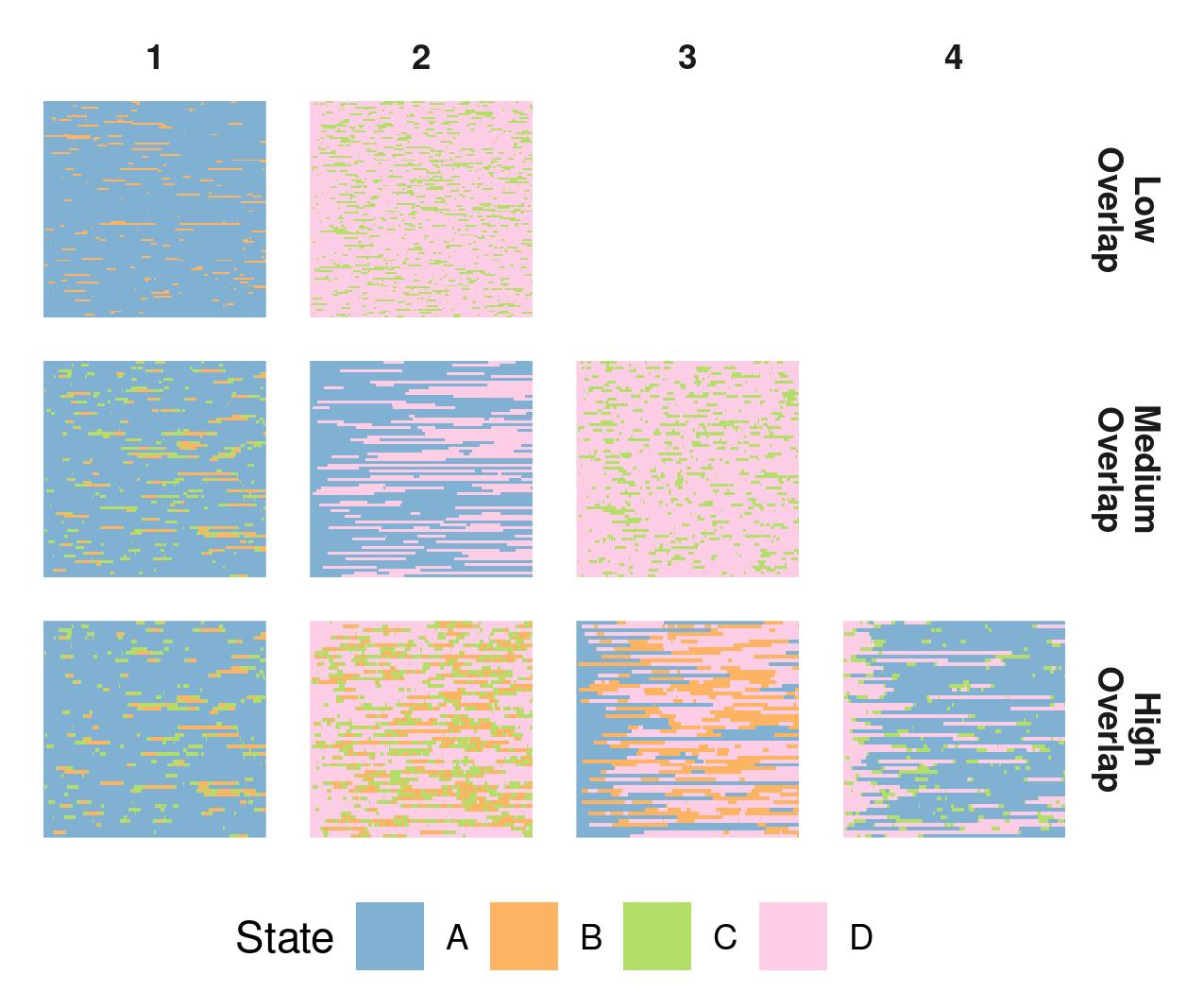}} 
    \subfloat[1 state duration varies]{\includegraphics[scale = .8]{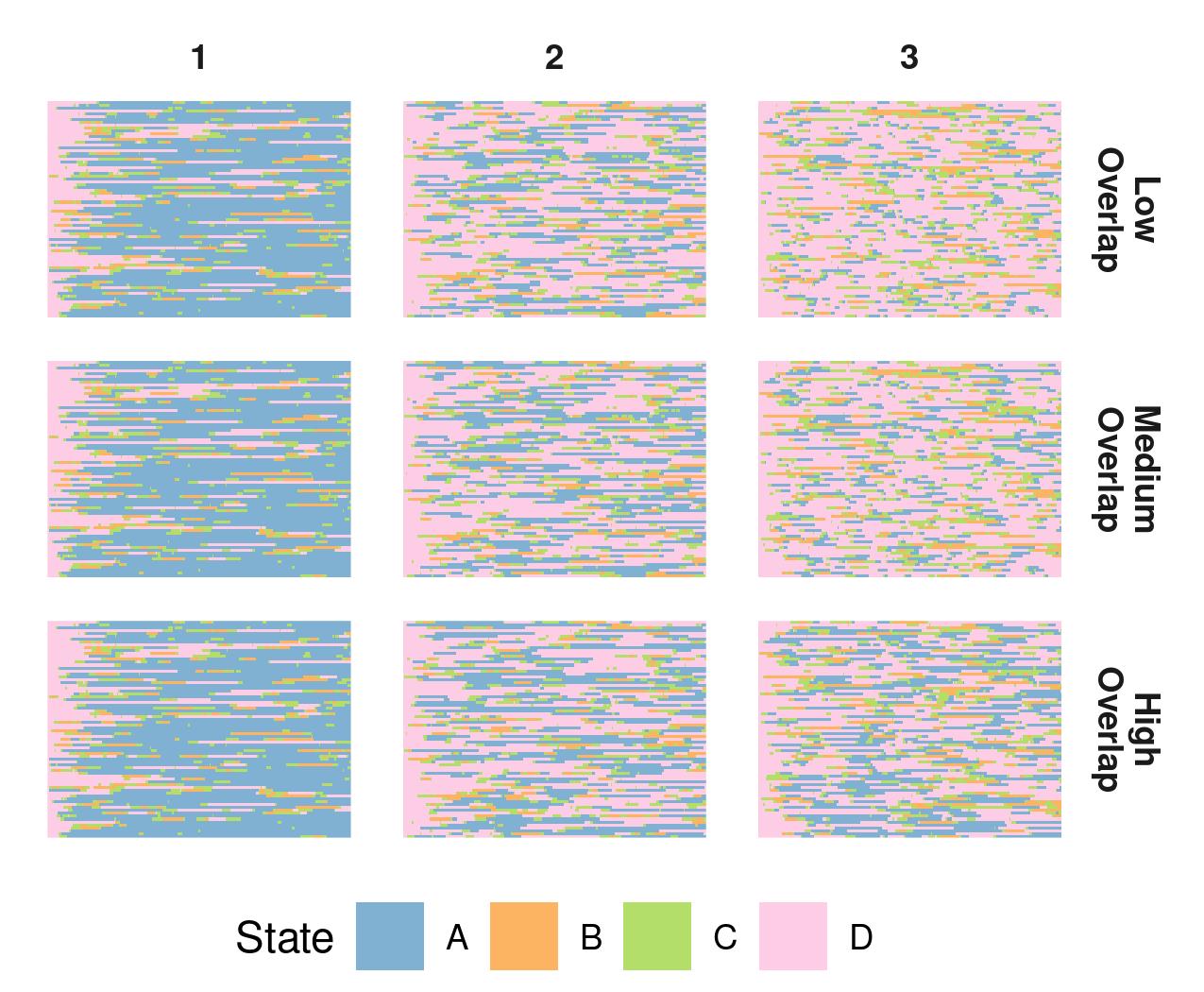}} \\
    \subfloat[2 state durations vary]{\includegraphics[scale = .8]{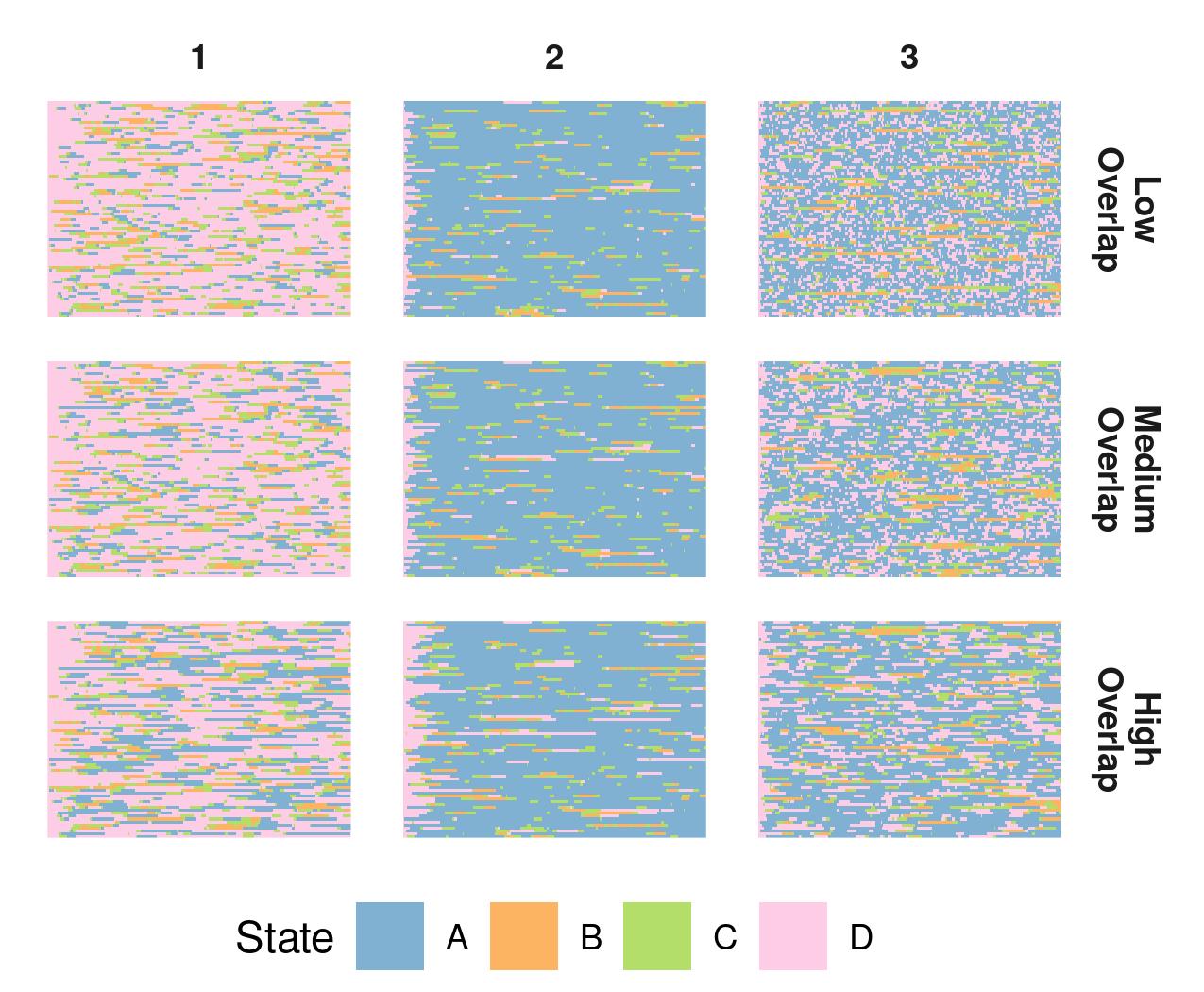}} 
    \subfloat[3 state durations vary]{\includegraphics[scale = .8]{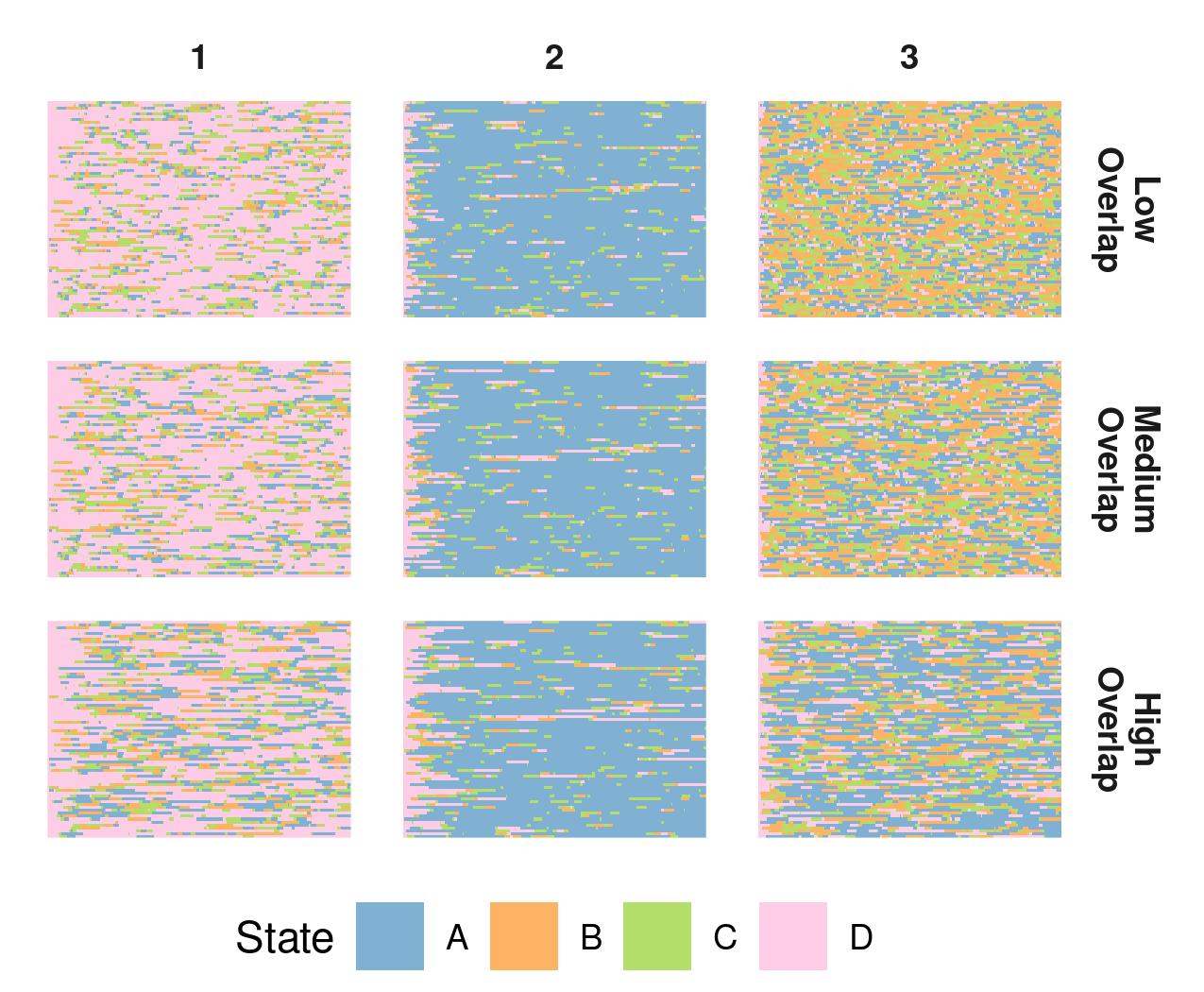}} 
    \caption{\small One sample set of simulated sequences from second-order Markov Chains (see Supplementary Table 1 for corresponding results). Columns differentiate each cluster and rows differentiate simulation scenarios by overlap level.}
    \label{fig:sup_sim_order2}
\end{figure}

\begin{figure}[bp]
    \centering
    \subfloat[States present vary ]{\includegraphics[scale = .8]{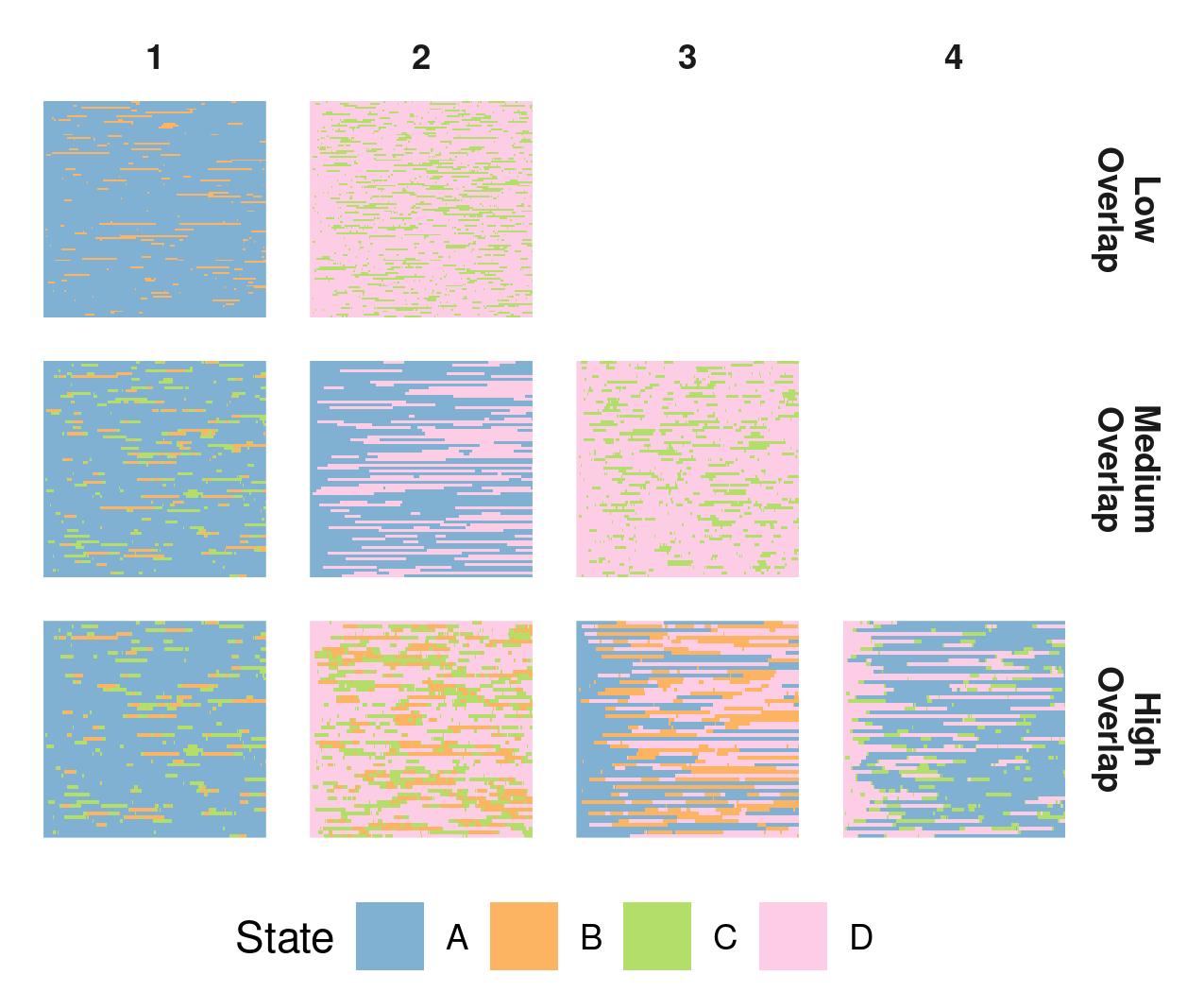}} 
    \subfloat[1 state duration varies]{\includegraphics[scale = .8]{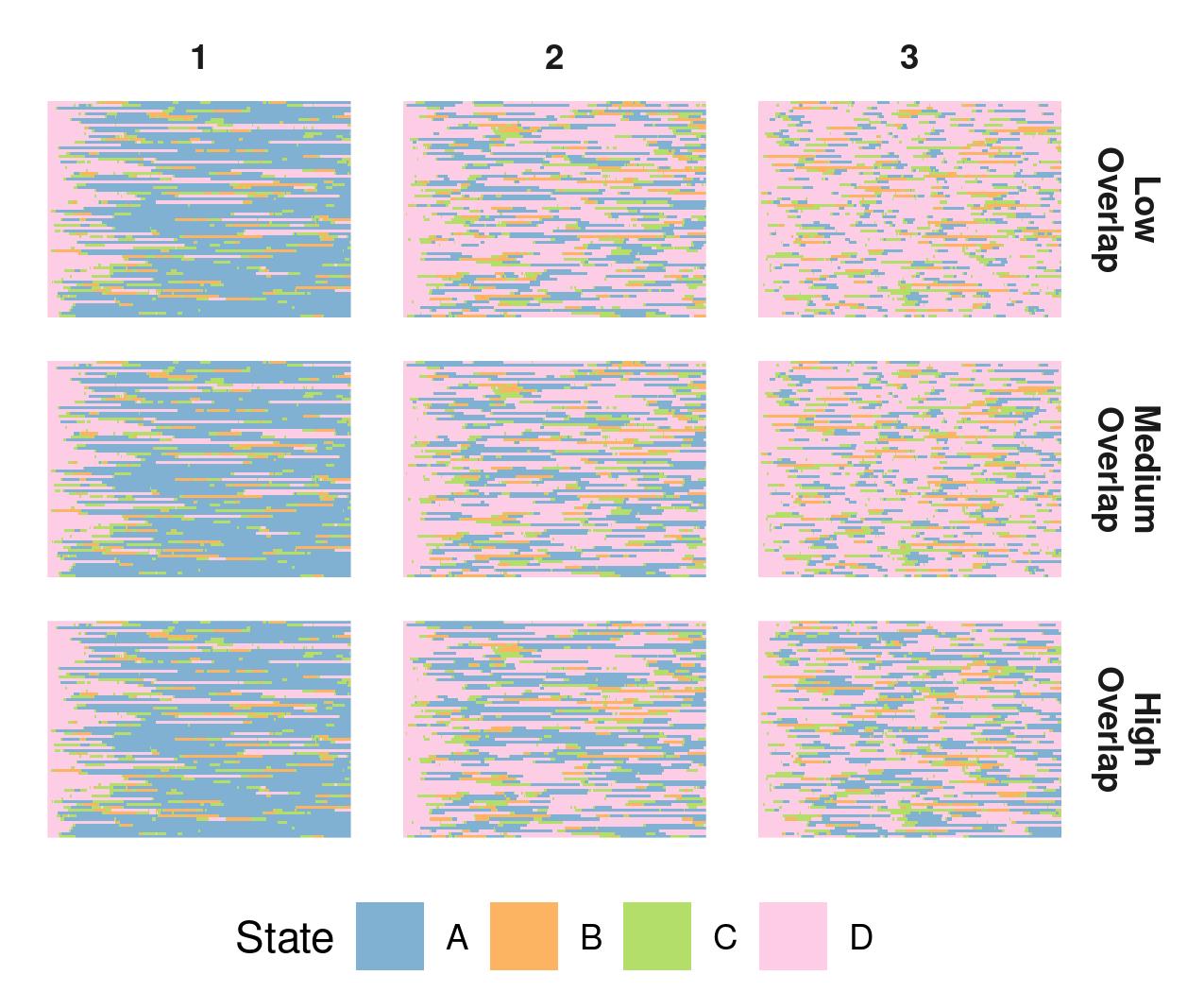}} \\
    \subfloat[2 state durations vary]{\includegraphics[scale = .8]{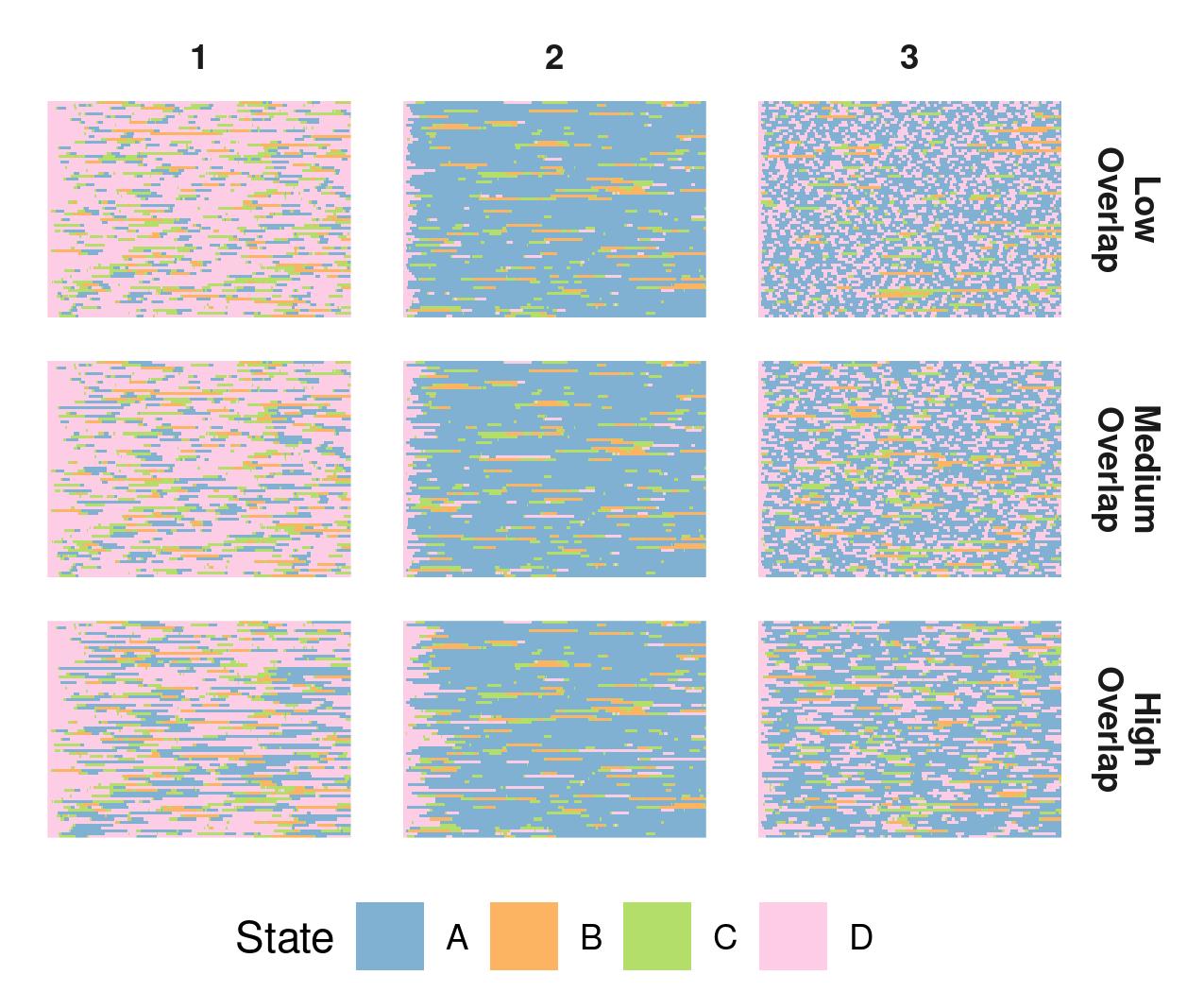}} 
    \subfloat[3 state durations vary]{\includegraphics[scale = .8]{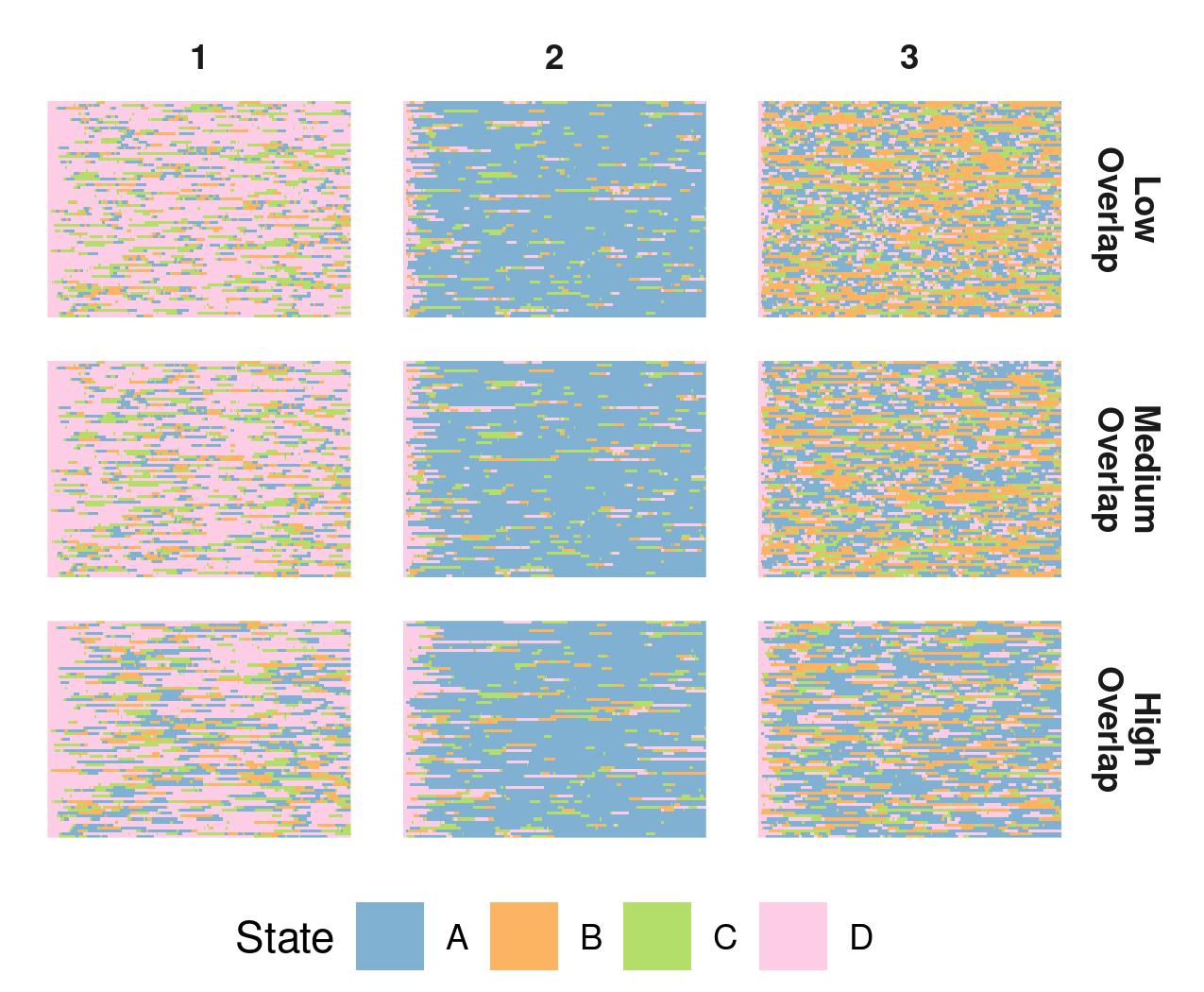}} 
    \caption{\small One sample set of simulated sequences from fifth-order Markov Chains (see Table 2 for corresponding results). Columns differentiate each cluster and rows differentiate simulation scenarios by overlap level.}
    \label{fig:sup_sim_order5}
\end{figure}

\begin{table}[h!]
\centering
\small
\caption{ \small Simulation results for sequences simulated by second-order Markov chains.}
\begin{tabular}{lllll|llll}
\hline
&& \multicolumn{3}{c}{\% Accuracy (SD)} & \multicolumn{4}{c}{No. Clusters} \\
Overlap & Type & 
 Hier.&  AMDC & nTreeClus 
  & True & 
  Hier.&  AMDC & nTreeClus \\
\hline
Low & State & 1.00 (0.00) & 1.00 (0.00) & 1.00 (0.00) & 2 & 2 & 2 & 2\\

Medium & & 0.73 (0.07) & 0.83 (0.03) & 0.79 (0.04) & 3 & 2 & 6 & 2\\

High & & 0.63 (0.06) & 0.66 (0.07) & 0.62 (0.05) & 4 & 2 & 2 & 2\\

Low & Dur. 1 & 0.62 (0.04) & 0.70 (0.05) & 0.68 (0.05) & 3 & 2 & 2 & 2\\

Medium & & 0.61 (0.04) & 0.68 (0.04) & 0.67 (0.04) & 3 & 2 & 2 & 2\\

High & & 0.51 (0.08) & 0.58 (0.03) & 0.57 (0.03) & 3 & 2 & 2 & 2\\

Low & Dur 2. & 0.69 (0.02) & 0.98 (0.01) & 0.97 (0.06) & 3 & 2 & 2 & 2\\

Medium & & 0.68 (0.02) & 0.95 (0.04) & 0.74 (0.07) & 3 & 2 & 2 & 2\\

High & & 0.60 (0.05) & 0.78 (0.04) & 0.61 (0.06) & 3 & 2 & 2 & 2\\

Low & Dur. 3 & 0.99 (0.01) & 1.00 (0.00) & 0.95 (0.03) & 3 & 2 & 3 & 3\\

Medium & & 0.92 (0.07) & 0.98 (0.01) & 0.84 (0.05) & 3 & 2 & 2 & 2\\

High & & 0.65 (0.03) & 0.81 (0.04) & 0.68 (0.04) & 3 & 2 & 2 & 2\\

\hline
\end{tabular}
\vspace{-0.1in}
\begin{flushleft}
\small Note: ``Overlap'' indicates the extent of the overlap in sequences between the true clusters (high overlap corresponds to less distinct clusters), ``Type'' indicates whether the true clusters differ by the states present or the average state duration.
\end{flushleft}
\label{table:sim_sec_res}
\end{table}

\begin{figure}[h!]
    \centering
    \includegraphics[scale = .17]{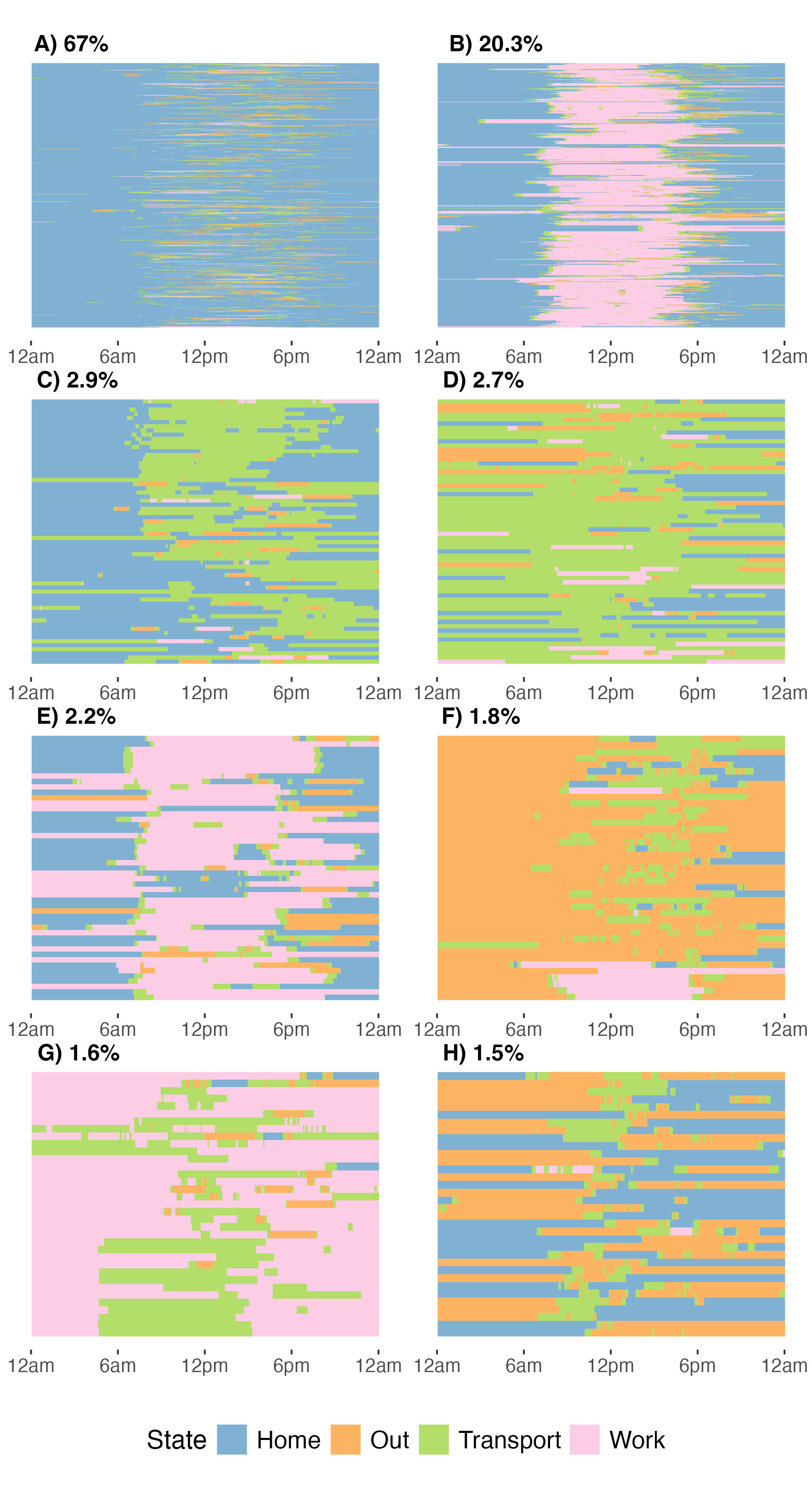}
    \caption{\small All optimal day sequence clusters from nTreeClus. Labels indicate the percent of all sequences in the respective cluster. Within each sub-figure, each row depicts a single sequence; row height has been standardized such that the sub-figure is the same height for all clusters.}
    \label{fig:ntrees_all}
\end{figure}

\begin{figure}[h!]
    \centering
    \includegraphics[scale = .17]{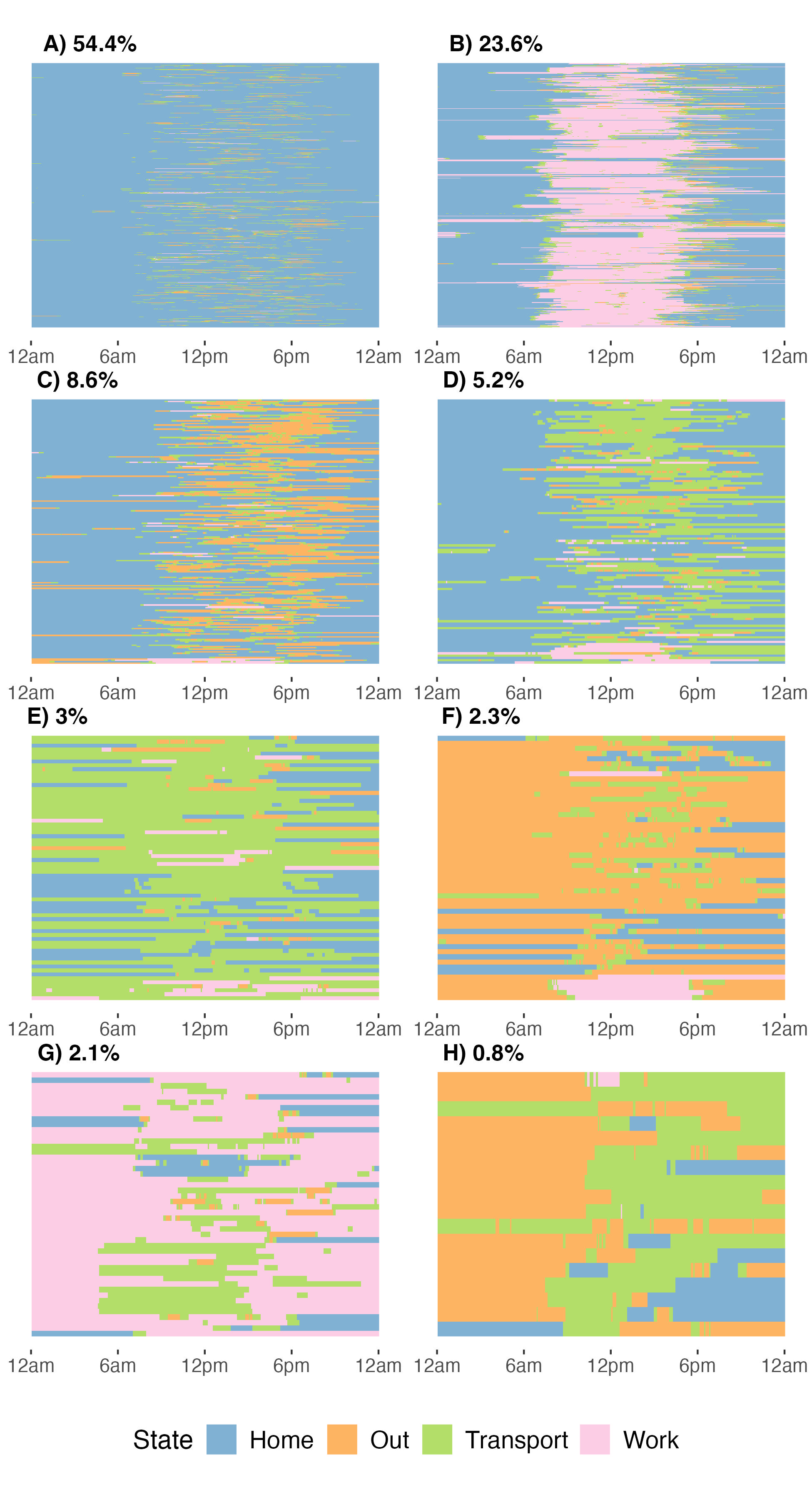}
    \caption{\small All optimal day sequence clusters from adjacency matrix decomposition clustering. Labels indicate the percent of all sequences in the respective cluster. Within each sub-figure, each row depicts a single sequence; row height has been standardized such that the sub-figure is the same height for all clusters.}
    \label{fig:amdc_all}
\end{figure}

\begin{figure}[h!]
    \centering
    \includegraphics[scale = .17]{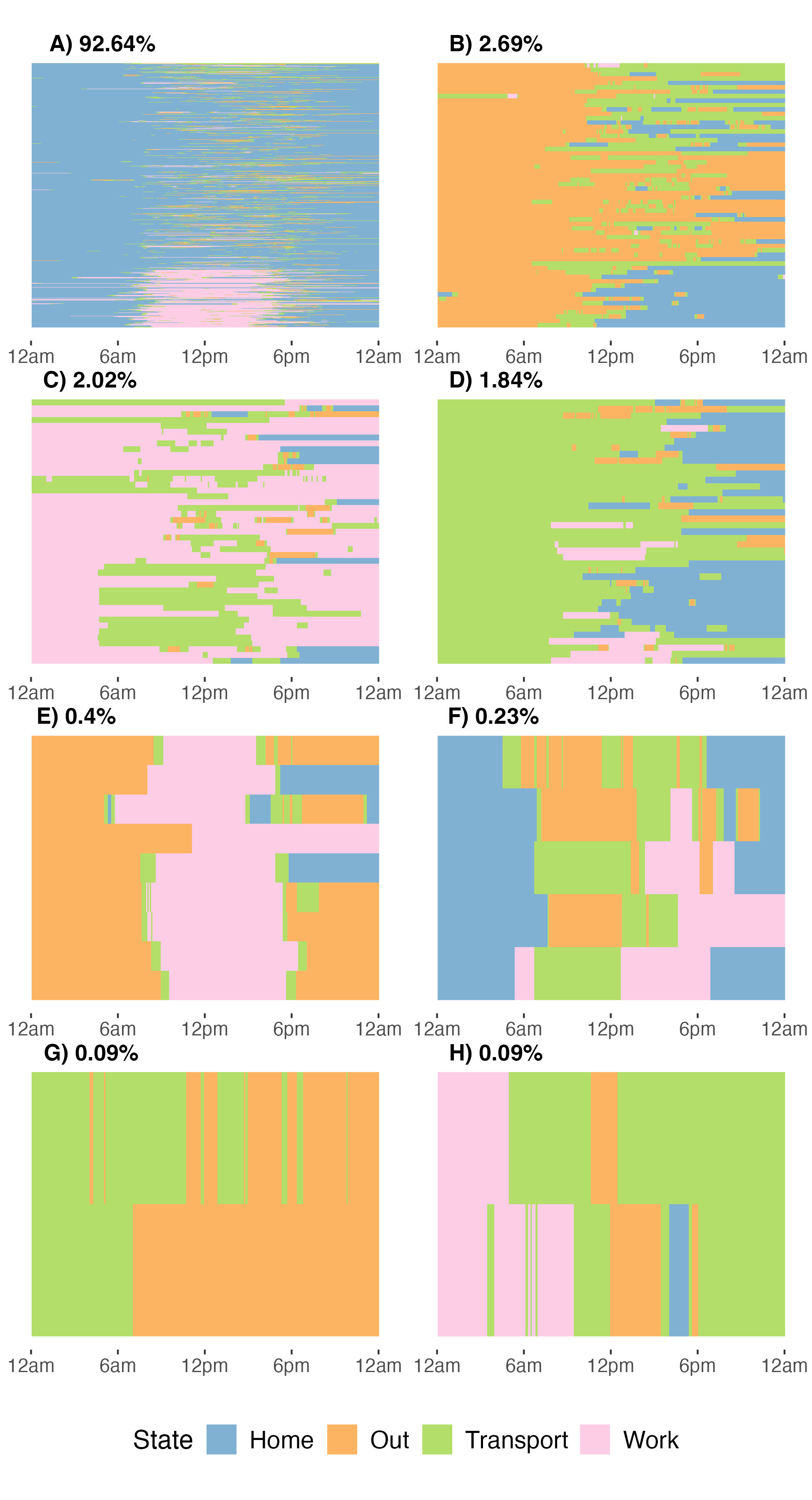}
    \caption{\footnotesize Day sequence clusters from  hierarchical clustering when selecting eight clusters (the optimal number of clusters from adjacency matrix decomposition clustering). Labels indicate the percent of all sequences in the respective cluster. Within each sub-figure, each row depicts a single sequence; row height has been standardized such that the sub-figure is the same height for all clusters.}
    \label{fig:hier_8_all}
\end{figure}

\begin{figure}[h!]
    \centering
    \includegraphics[scale = .21]{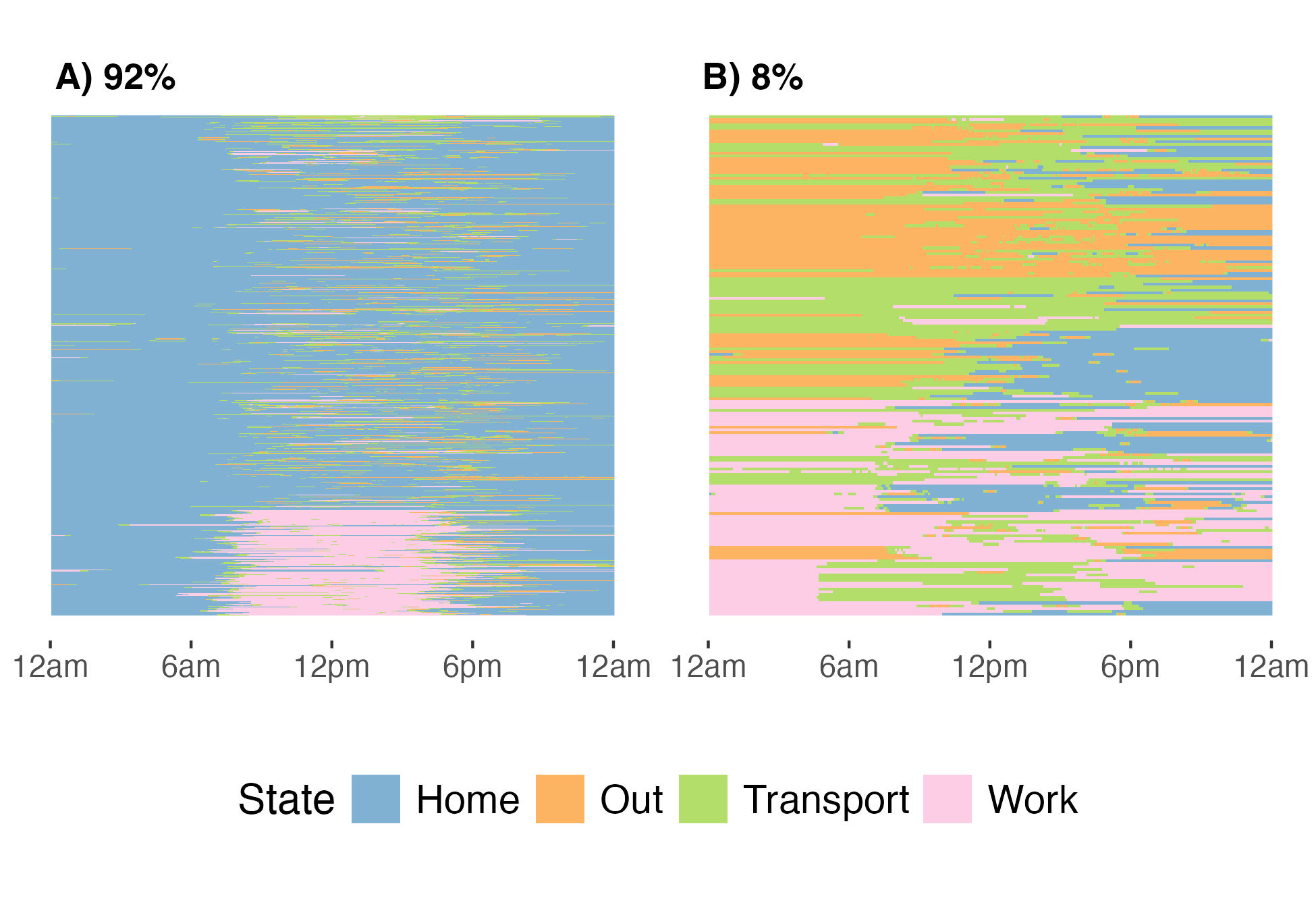}
    \caption{\small All day sequence clusters from the nTreeClus method where state position is incorporated. Labels indicate the percent of all sequences in the respective cluster. Within each sub-figure, each row depicts a single sequence; row height has been standardized such that the sub-figure is the same height for all clusters.}
    \label{fig:ntrees_all_pos}
\end{figure}

\begin{table}[h!]

\centering
\centering
\caption{Fixed effects from the mixed effects logistic regression of COVID-19 concern regressed on adjacency matrix decomposition cluster assignment and random intercepts for individuals. Cluster assignments are shown in Supplementary Figure 5. In this model, cluster A is the reference group.}
\begin{tabular}{l|llll}
\hline
  & Estimate & OR    & SE   & p-value             \\ \hline 
Intercept & -2.59    & 0.08  & 0.24 &                \\
B        & 0.81     & 2.25  & 0.24 & \textless 0.01* \\
C        &  1.31     & 3.70  & 0.26 & \textless 0.01* \\
D        & 1.01     & 2.75  & 0.41 & 0.01*           \\
E        & 1.33     & 3.76  & 0.48 & 0.01*           \\
F         & 1.30     & 3.66  & 0.50 & 0.01*           \\
G         & 0.74     & 2.10  & 0.60 & 0.22           \\
H         & 2.83     & 16.96 & 0.96 & \textless 0.01* \\
\hline
\end{tabular}
\end{table}

\begin{figure}[h!]
    \centering
    \includegraphics[scale = .17]{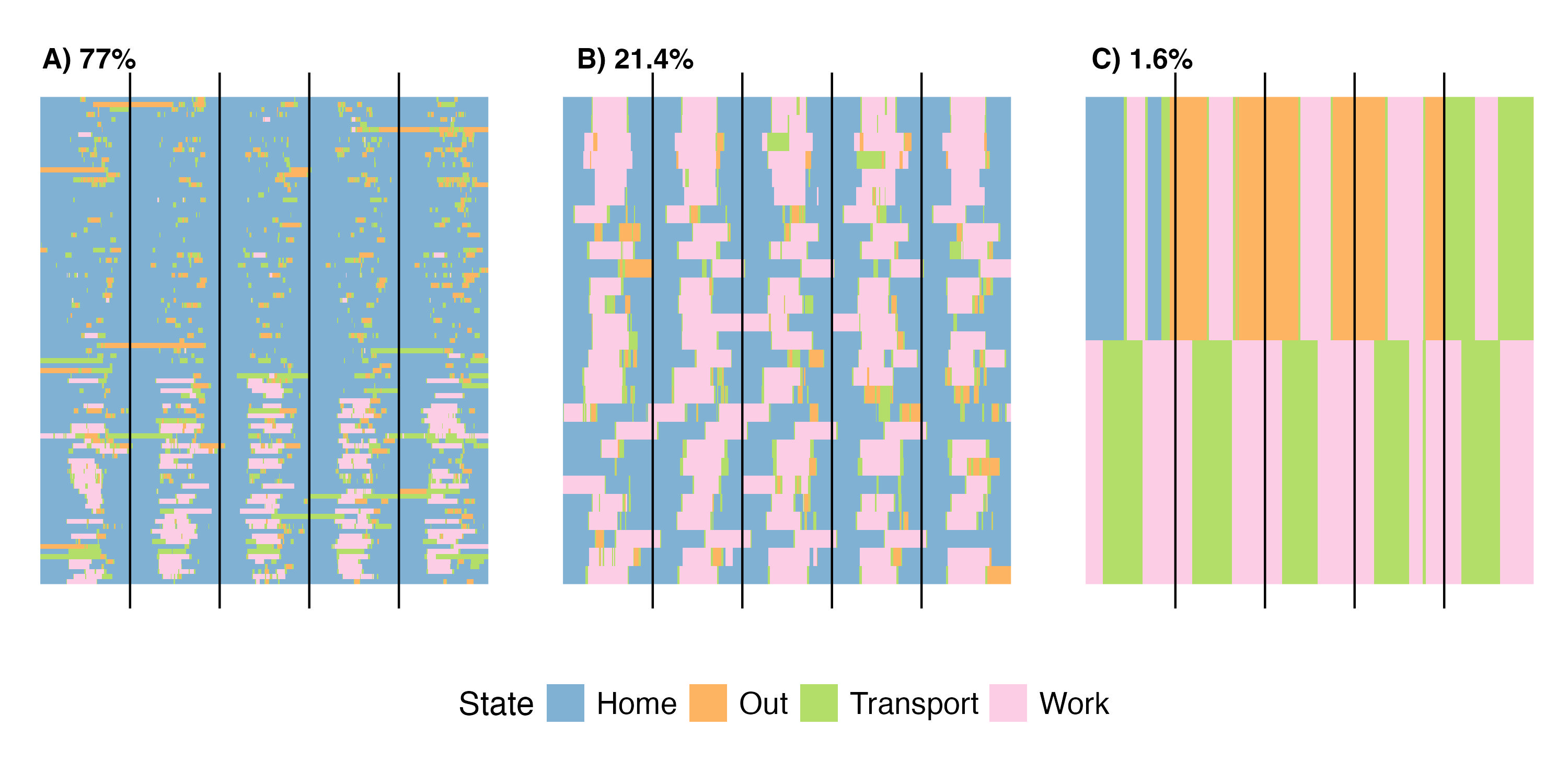}
    \caption{\small All optimal week sequence clusters from nTreeClus. Black vertical lines separate each of the five days. Labels indicate the percent of all sequences in the respective cluster. Within each sub-figure, each row depicts a single sequence; row height has been standardized such that the sub-figure is the same height for all clusters.}
    \label{fig:ntrees_week_all}
\end{figure}

\begin{figure}[h!]
    \centering
    \includegraphics[scale = .17]{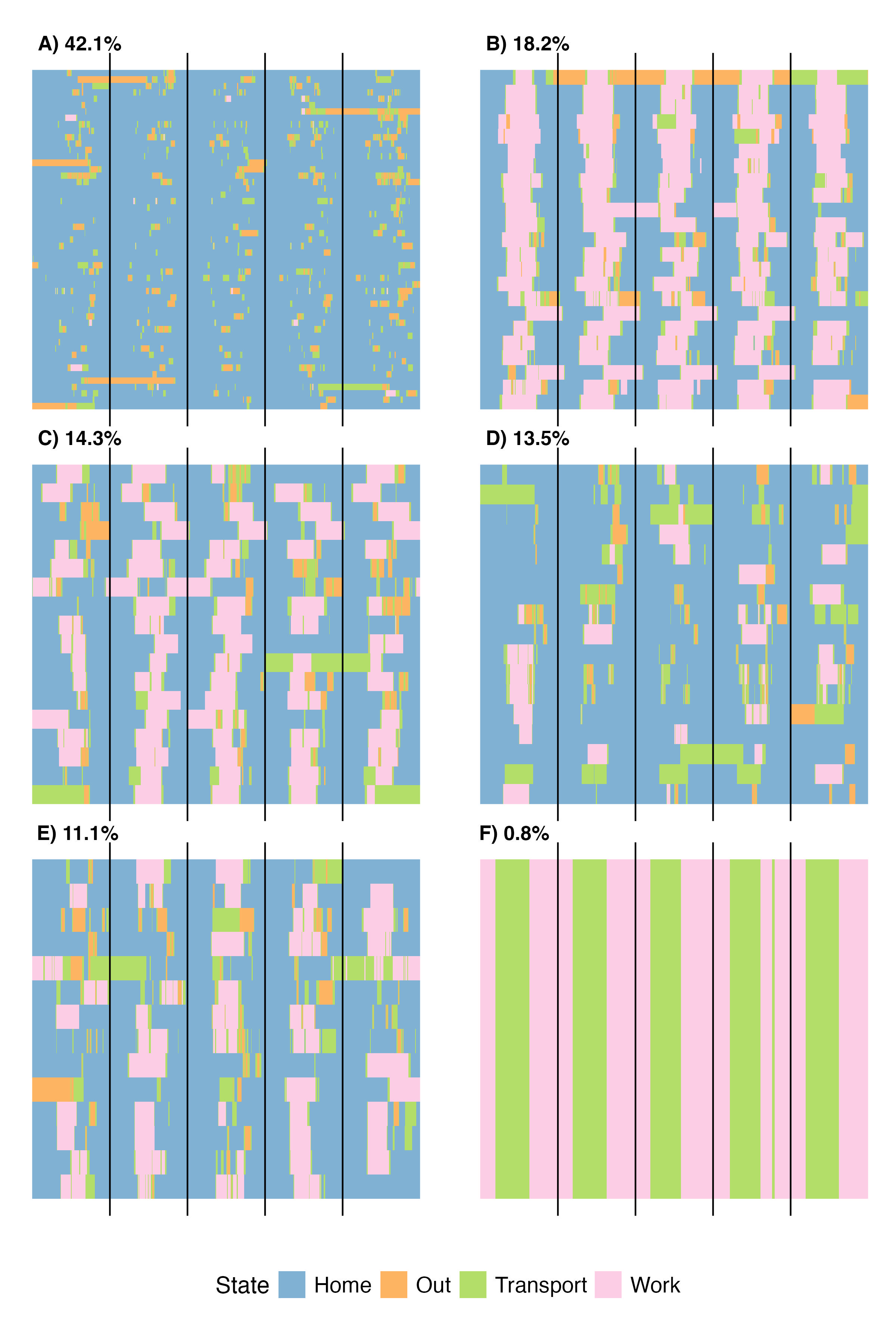}
    \caption{\footnotesize All optimal week sequence clusters from adjacency matrix decomposition clustering. Black vertical lines separate each of the five days. Labels indicate the percent of all sequences in the respective cluster. Within each sub-figure, each row depicts a single sequence; row height has been standardized such that the sub-figure is the same height for all clusters.}
    \label{fig:amdc_week_all}
\end{figure}

\begin{figure}[h!]
    \centering
    \includegraphics[scale = .17]{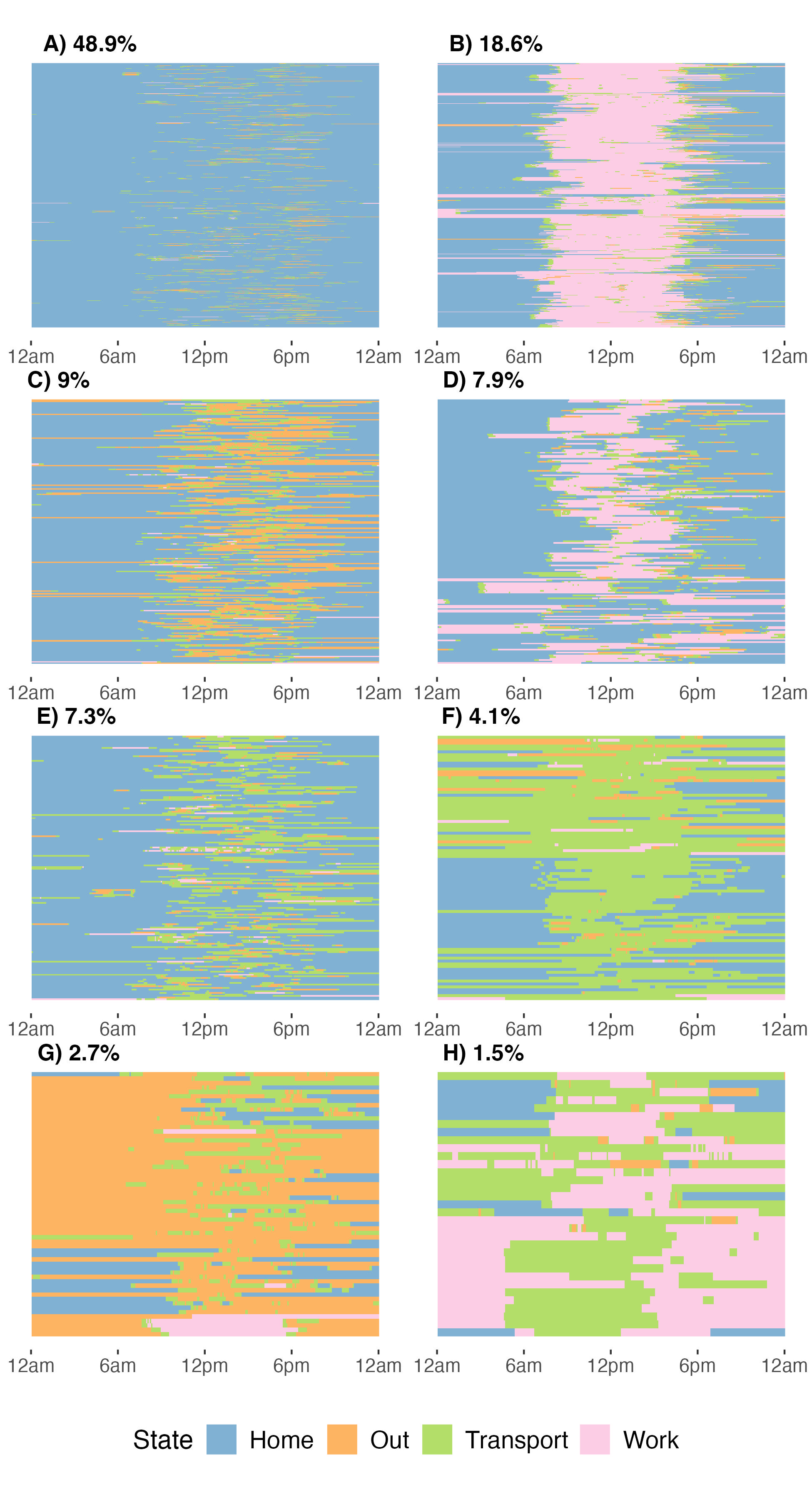}
    \caption{\footnotesize All optimal 9AM-5PM weighted day sequence clusters (relative weight of 2) from adjacency matrix decomposition clustering. Labels indicate the percent of all sequences in the respective cluster. Within each sub-figure, each row depicts a single sequence; row height has been standardized such that the sub-figure is the same height for all clusters.}
    \label{fig:amdc_9_5_all}
\end{figure}

\begin{figure}[h!]
    \centering
    \includegraphics[scale = .17]{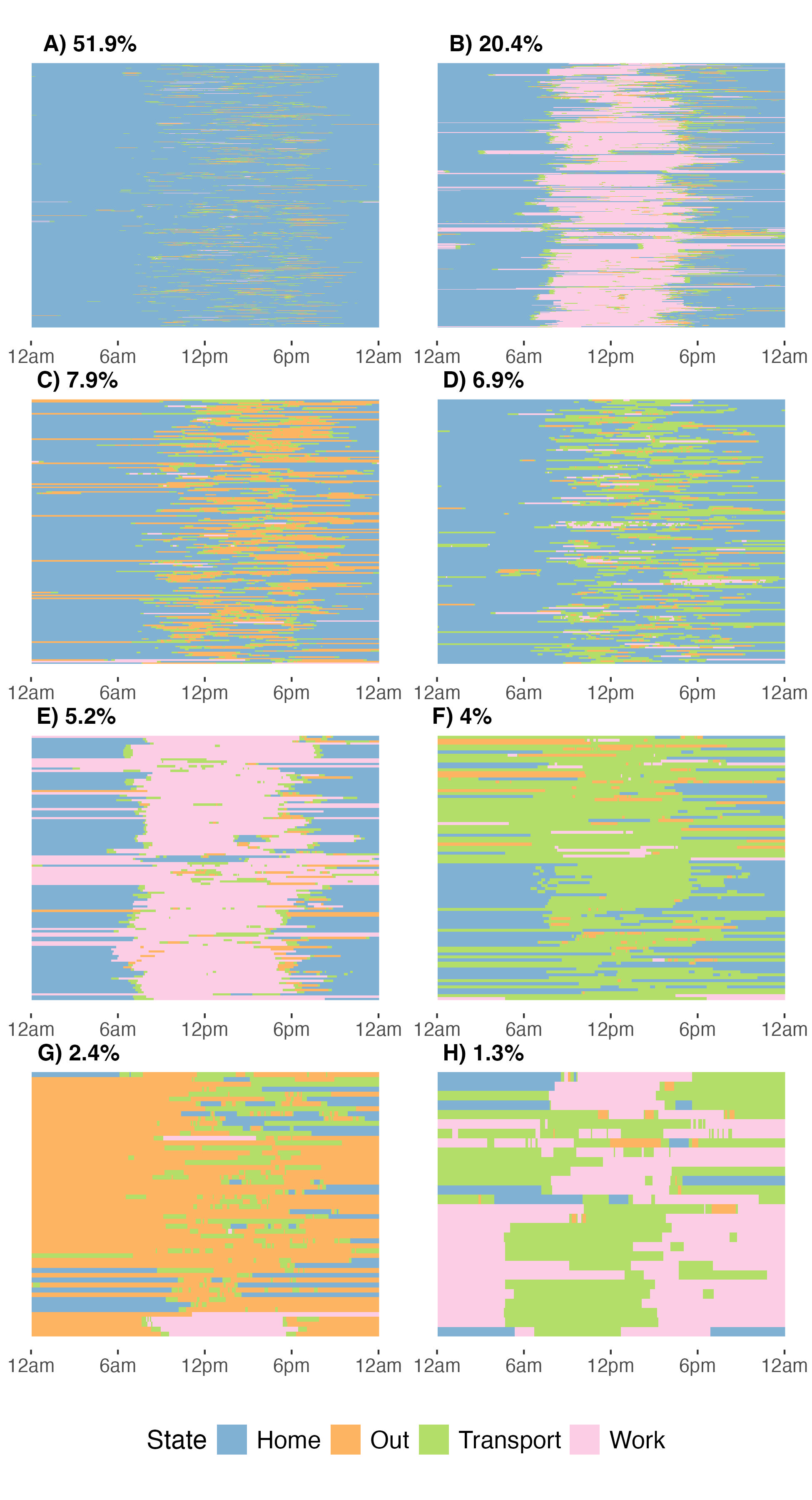}
    \caption{\footnotesize All optimal 9AM-5PM weighted day sequence clusters (relative weight of 1.5) from adjacency matrix decomposition clustering. Labels indicate the percent of all sequences in the respective cluster. Within each sub-figure, each row depicts a single sequence; row height has been standardized such that the sub-figure is the same height for all clusters.}
    \label{fig:amdc_9_5_w15}
\end{figure}

\begin{figure}[h!]
    \centering
    \includegraphics[scale = .17]{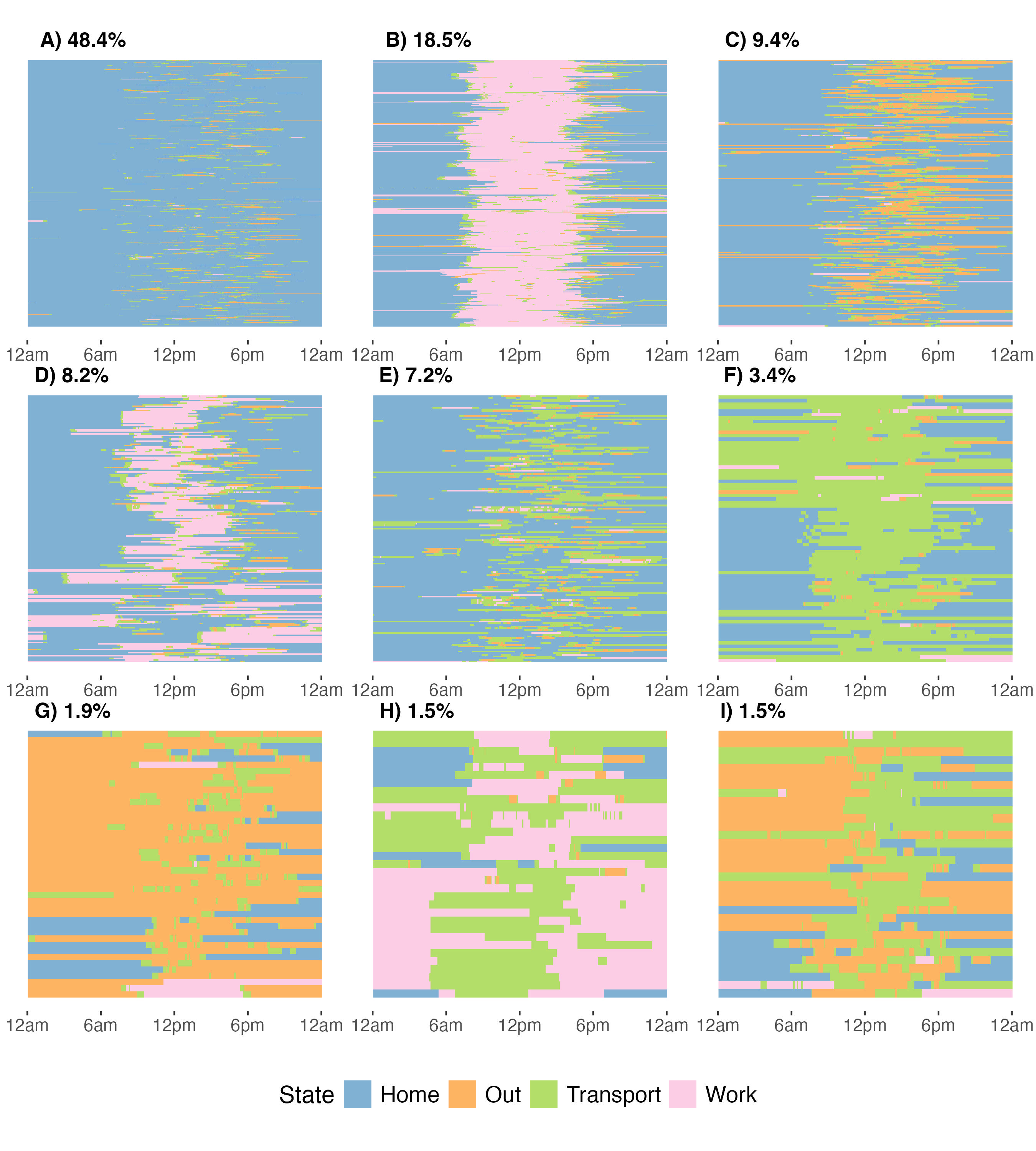}
    \caption{\small All optimal 9AM-5PM weighted day sequence clusters (relative weight of 2.5) from adjacency matrix decomposition clustering. Labels indicate the percent of all sequences in the respective cluster. Within each sub-figure, each row depicts a single sequence; row height has been standardized such that the sub-figure is the same height for all clusters.}
    \label{fig:amdc_9_5_w25}
\end{figure}

\begin{figure}
    \centering
    \includegraphics[width=0.65\linewidth]{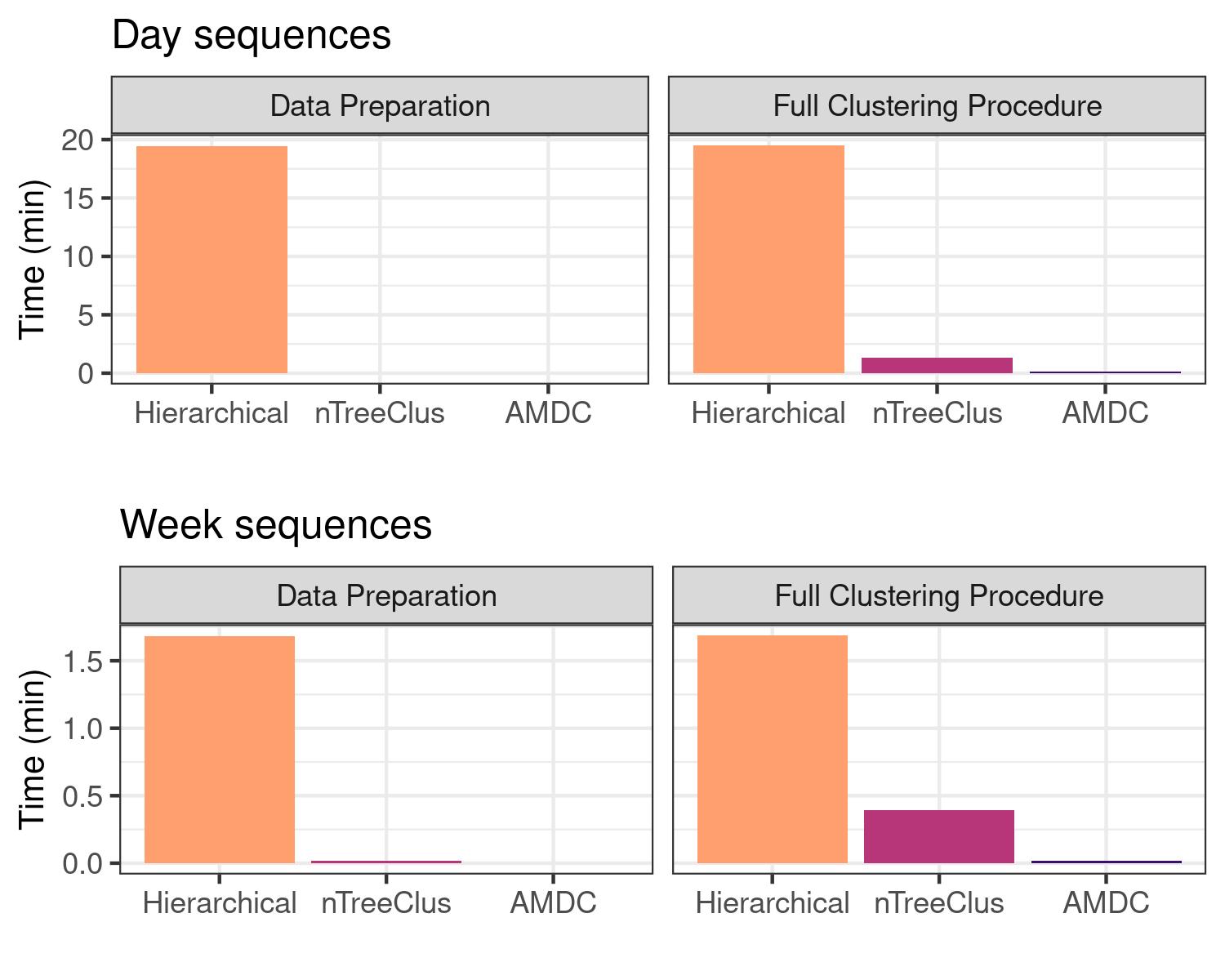}
    \caption{ \small Computational time for each method for day and week sequences. Note that these times are specific to the given machine and code structure used to perform the methods.}
    \label{fig:enter-label}
\end{figure}

\pagebreak

\subsection*{1 Stability Analysis}
We implement the stability procedure proposed by Yu et al., 2019. To account for the correlation within individuals, we first sample individuals with replacement and then resample day sequences from the sampled individuals. We perform the clustering method on each bootstrapped dataset; the resulting cluster centers are then used to obtain clusters of the original data. Thus, each bootstrapped dataset results in clusters of the original non-bootstrapped dataset. This allows us to calculate stability for each observation in the original data. In this procedure, stability is scored from zero to one (one being perfectly stable) for each observation; overall and cluster-level stability is calculated by taking the mean or median of the stability values for the observations.

The clusters from adjacency matrix decomposition clustering have mean and median stability of $0.79$ and $0.88$ across all observations. Clusters A and B (Figure 1, Panel 2) are the most stable with a mean stability of $0.87$ and $0.81$ respectively (median: $0.92$ and $0.87$). Cluster H (Supplementary Figure 5) is not stable due to the small number of sequences within the cluster. All remaining clusters have mean stability ranging from $0.51 - 0.67$ and median stability ranging from  $0.54 - 0.74$. We also found that the optimal number of clusters selected is stable; in 98.2 \% of the bootstrapped datasets, $7-9$ optimal clusters are selected. 
\end{document}